\documentclass[aps,pra,twocolumn,a4paper,showpacs,superscriptaddress,floatfix,10pt]{revtex4-1}
\pdfoutput=1

\usepackage{amsfonts}
\usepackage{amssymb}
\usepackage{amsmath}
\usepackage{amsthm}
\usepackage{bm}
\usepackage{booktabs}
\usepackage{complexity}
\usepackage{dcolumn}
\usepackage[pdftex]{graphicx,color}
\usepackage{pdfsync}
\usepackage{rotate}
\usepackage{siunitx}
\usepackage{subfigure}
\usepackage{framed}

\usepackage[pdftex]{hyperref}

\let\oldmarginpar\marginpar
\renewcommand\marginpar[1]{\-\oldmarginpar[\raggedleft\tiny #1]%
{\raggedright\tiny #1}}

\DeclareMathOperator{\Tr}{Tr}

\newcommand{\avg}[1]{\left< #1 \right>}
\newcommand{\bra}[1]{\langle#1|}
\newcommand{\ket}[1]{|#1\rangle}
\newcommand{\braket}[2]{\langle#1|#2\rangle}

\newcommand{\ie}{{\it i.e. }}

\newcommand{\eg}{{\it e.g. }}

\def\mean{\mathrm{mean}}
\def\var{\operatorname{Var}}

\graphicspath{{figs/}}

\begin{document}

\title{On the quantum spin glass transition on the Bethe lattice}

\author{G.Mossi}
\affiliation{SISSA, via Bonomea 265, 34136 Trieste, Italy}
\affiliation{INFN, Sezione di Trieste, Via Valerio 2, 34127 Trieste, Italy}

\author{T.Parolini}
\affiliation{Universit\`a di Milano, via Celoria 16, 20122 Milan, Italy}

\author{S.Pilati}
\affiliation{Abdus Salam ICTP, Strada Costiera 11, 34151 Trieste, Italy}

\author{A. Scardicchio}
\affiliation{Abdus Salam ICTP, Strada Costiera 11, 34151 Trieste, Italy}
\affiliation{INFN, Sezione di Trieste, Via Valerio 2, 34127 Trieste, Italy}
\date{June 21, 2016}

\begin{abstract}
We investigate the ground-state properties of a disorderd Ising model with uniform transverse field on the Bethe lattice, focusing on the quantum phase transition from a paramagnetic to a glassy phase that is induced by reducing the intensity of the transverse field. We use a combination of quantum Monte Carlo algorithms and exact diagonalization to compute R\'enyi entropies, quantum Fisher information, correlation functions and order parameter. We locate the transition by means of the peak of the R\'enyi entropy and we find agreement with the transition point estimated from the emergence of finite values of the Edwards--Anderson order parameter and from the peak of the correlation length. We interpret the results by means of a mean-field theory in which quantum fluctuations are treated as massive particles hopping on the interaction graph. We see that the particles are delocalized at the transition, a fact that points towards the existence of possibly another transition deep in the glassy phase where these particles localize, therefore leading to a many-body localized phase.

\end{abstract}

\maketitle

\section{Introduction}
Recent technological advances in the pursuit of building an adiabatic quantum computer \cite{farhi2001quantum,kadowaki98quantum,boixo2014evidence,laumann2015quantum} have renewed the interest of the community in isolated, classical spin systems which are given a quantum dynamics by means of a transverse, uniform field. Among these, spin glasses have received a lot of attention, mostly due to their connections with difficult (NP-complete) computational problems \cite{kirkpatrick1983optimization,kirkpatrick1994critical,monasson1999determining,mezard2002analytic}. In this paper we report on the investigations of such transverse-field quantum spin glasses on a regular random graph, or Bethe lattice, by focusing on the entanglement properties and correlation functions of the ground state. 

\noindent Models of spin glasses with non-commuting terms (\emph{quantum spin glasses} for short) are not new in the literature and have been analyzed on several geometries to various degrees of details and, although no exact solution exists, a lot is known \cite{bray1980replica,goldschmidt1990ising,buccheri2011structure, andreanov2012long} including connections with quantum complexity theory \cite{laumann2010random,laumann2010product}. 

\noindent As it seems to be the rule in these problems, the geometry in which one can expect to do more progress is the fully-connected graph analog to the Sherrington-Kirkpatrick model \cite{sherrington1975solvable,thouless1977solution,parisi1980order}, which however has a limited concept of locality, since any couple of spins is at minimum distance. The second in line are spin models on the Bethe lattice where the Hamiltonian is local and some kind of iteration procedure can be used to solve them \cite{thouless1986spin,mezard2001bethe}. Some of the concepts developed in the replica theory of spin glasses, chiefly, the replica symmetry breaking cavity method of Ref. \cite{mezard2001bethe} have been generalized to transverse field Ising spin glasses \cite{laumann2008cavity}, quantum ferromagnets \cite{krzakala2008path}, and combinatorial optimization problems \cite{farhi2012performance}, allowing one to  
see certain features of the transition from the paramagnetic to the spin-glass phase which occurs when the transverse-field intensity is reduced, directly in the thermodynamic limit. However this quantum generalization of the cavity method has not yet turned into a useful tool to study purely quantum aspects of the phases, like the entanglement entropy.

Another reason that motivates our research, which only at first sight might seem detached from the previous discussion, are the recent developments in the theory of disordered, isolated quantum systems, in particular the study of their unitary dynamics (as opposed to the partition function), which comes under the umbrella of many-body localization (MBL) \cite{basko2006metal, pal2010mb, nandkishore2014many}. This body of work suggests that for disordered quantum systems a phase of non-ergodic \cite{de2013ergodicity,laumann2014many,luitz2015many,goold2015total,lerose2015coexistence} or even integrable dynamics \cite{huse2014phenomenology,imbrie2014many,ros2015integrals,chandran2015constructing} exists which prevents a statistical mechanics description of the equilibrium properties of the system. This phenomenon has been pointed out as a possible bottleneck for the adiabatic algorithm \cite{altshuler2010anderson} solving a classically difficult problem, but these claims have been criticized and are currently under investigation \cite{knysh2010relevance,laumann2015quantum}. The issue, in brief, is that small gaps are natural in a many-body phase due to the above-mentioned emergent integrability. However, a quantitative analysis of this smallness is based on numerical investigations and is still incomplete; a general theoretical framework is still missing.

Therefore one is naturally lead to ask whether there is an MBL phase inside the glassy phase or, perchance, it spans the entire glassy phase (a similar observation has been made in \cite{laumann2014many,baldwin2016many}). An MBL phase is a region of parameter space in which the Langevin forces disappear at a \emph{microscopic} level due to quantum fluctuations, inhibiting transport of all conserved quantities (but not the development of entanglement among far away spins \cite{vznidarivc2008many,bardarson2012unbounded}); in a spin glass phase Langevin forces still exists, as it is testified by the system equilibrating in a (local) minimum of the free energy, although (in particular in high-dimensional lattices like our RRG) such forces might not be sufficiently strong to overcome the barriers that are formed in between local minima and ergodicity is broken at a \emph{macroscopic} level. Therefore there is no logical reason why the two transitions should coincide. 

The main goal of this paper is to identify and characterize the quantum phase transition (QPT) from the paramagnetic to the glassy phase. We use exact diagonalization and quantum Monte Carlo algorithms to study large system sizes. We locate the critical point of the transition with high precision using the position of the peak of the R\'enyi-2 entanglement entropy and the emergence of finite values for the Edwards--Anderson order parameter, finding identical results. We see that the critical correlation lengths converge to a finite value in the thermodynamic limit. The analysis of the quantum Fisher information, which is found to be nonextensive, suggests that multipartite entanglement never diverges but is limited to pairs of spins for any system size.

We also give some indications that at the quantum spin glass transition there is no sign of many-body localization by presenting a mean-field picture of the transition in which delocalized quantum excitations are present and showing that it gives a reasonable prediction of the transition point and of the observed properties of the ground state wave function. The picture that emerges is that this transition does not bear any sign of MBL and therefore we conjecture that, if MBL exists in this model, it will be inside the glassy phase. We will comment on the implications of this result for present attempts to adiabatic quantum computation.

The paper is organized as follows: in Section \ref{sect:introduction} we review the known results for both the classical and quantum transition, as well as the motivations that underlie this work; in Section \ref{sect:numerical} we describe the numerical methods we employ and we present the results for the order parameter, the connected correlation functions, the quantum Fisher information and the R\'enyi entanglement entropy of the ground state; in Section \ref{sect:mean_field} we present a mean-field theory which explains the above observations; we present the conclusions and open directions for further work in Section \ref{sect:conclusions}. 

\section{The model and known results}\label{sect:introduction}

We study the model of a spin glass on a regular random graph (RRG) with $N$ vertices, with transverse field:
\begin{equation}
\label{eq:tfisg}
H=-\sum_{\langle i,j \rangle}J_{ij}\sigma^z_i\sigma^z_j-\Gamma \sum_{i=1}^N \sigma_i^x,
\end{equation}
where $\sigma_i^{a}$ (for $a = x,z$) is a Pauli matrix acting on the $i$-th spin of the system. The couplings $J_{ij}=\pm J$ with probability $1/2$ and the RRG has fixed connectivity $K+1$. $K$ is then called the branching number of the associated Bethe lattice. For convenience we take $K=2$ in the rest of the paper, in particular in the numerical treatments, since this allows us to have systems of larger ``linear dimension'', the diameter of the RRG $L\simeq \ln N/\ln K$. 

\begin{figure}
\begin{center}
  \includegraphics[width=.89\linewidth]{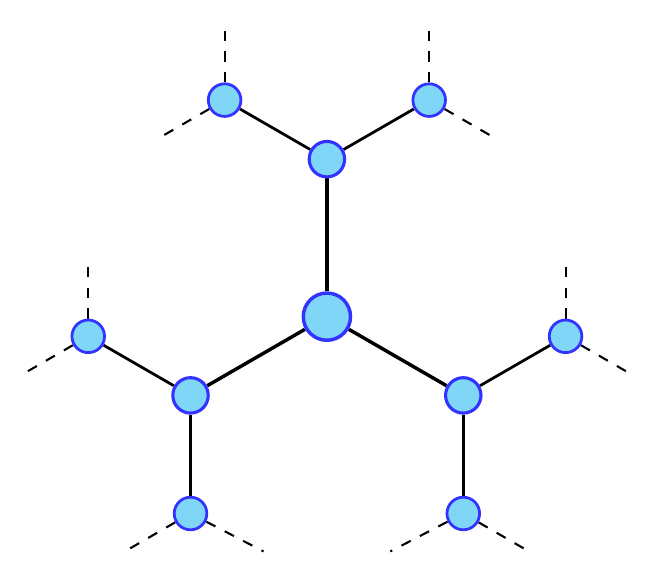}
\caption{The Bethe lattice of branching number $K$ is the unique $(K + 1)$-regular tree on an infinite number of nodes, depicted above for $K = 2$. Starting from any point in the graph, the $r$-th ``shell'' around the point has $(K + 1)K^{r-1}$ elements, which means that in the truncated tree only a fraction $\sim 1/K$ of the spins belongs in the bulk. In order to avoid dealing with the boundary, we will think of the Bethe lattice as the local large size limit of random regular graphs.}
  \label{fig:qea_pure}
\end{center}
\end{figure}

We consider the partition function of the system ($\beta=1/T$)
\begin{equation}
Z=\Tr e^{-\beta H},
\end{equation}
assuming the dynamics is ergodic. This is almost certainly true if the system is coupled to a bath (which is the situation we consider here) but it is also probably true for an isolated system in a large region of parameters, including the paramagnetic region, although a complete analysis of this problem would require an analysis on the line of Refs \cite{laumann2014many,baldwin2016many}.
In particular we assume that taking the limit $T\to 0$ the region of parameters investigated is within the ergodic region (this is one of the reasons we do not push our analysis deep in the spin glass phase).

The transition at $\Gamma=0$ can be called the \emph{classical} transition as it is due solely to the Langevin forces generated by the interaction with the bath. This transitions is supposed to determine the characteristics of the whole line extending up to, but not including, $T=0$. This can be understood as, upon Suzuki--Trotter, the system can be given a transverse dimension of size $\beta=1/T$, or more precisely $\sim J/T$. When one is close to the transition such that the critical system size exceeds this length the system can be ``renormalized'' back to the original RRG. This cannot be done when $T=0$.

We now discuss the classical transition, which has been studied extensively in the literature, and the known computational complexity considerations for the model at hand. 

\subsection{Classical transition}
So we are presented with a phase diagram in the $(T,\Gamma)$-plane, where the classical spin glass transition ($\Gamma=0$) occurs at
\begin{equation}
\label{eq:TcSG}
T_c=J/\tanh^{-1}(1/\sqrt{K}),
\end{equation}
while the location of the quantum spin glass transition $\Gamma_c$ at $T=0$ is not known analytically and a convex line joins these two points.

For $K=2$ this is (setting the units of energy such that $J=1$) $T_c\simeq 1.135$ or $\beta_c=0.881$, and $\Gamma_c$ between $1.5$ and $2.0$ from the analysis of Ref.\cite{laumann2008cavity}. It is also important to notice that for large $K$, $T_c\simeq J\sqrt{K}.$ We will see that a mean field theory for the quantum fluctuations at $T=0$ also predicts $\Gamma_c\propto J\sqrt{K}$ which means that $J_{ij}'s$ need to be scaled with $1/\sqrt{K}$ for large $K$ although the origin of the two scalings is apparently quite different.

It is convenient now to review some known properties of the classical transition. First of all, we recall that the system can be solved numerically, directly in the thermodynamic limit, to arbitrary degrees of accuracy with the cavity method of \cite{mezard2001bethe}. This method introduces the cavity fields $h_i$ which are the effective fields acting on a given spin, once one of the $K+1$-th link has been removed.

It is possible to find a recursion equation for the cavity fields of the spin 0 once those of the spins $1,\dotsc,K$ are known
\begin{equation}
\label{eq:cavityeq}
h_0=\frac{1}{\beta}\sum_{i=1}^K\tanh^{-1}(\tanh(\beta J_{0i})\tanh(\beta h_i)).
\end{equation}
These fields are random i.i.d.\ (since the correlations are negligible for large system sizes) and distributed according to a probability distribution which is stable under (\ref{eq:cavityeq}). The distribution which is stable under iterations in the paramagnetic phase is
\begin{equation}
\label{eq:Ppara}
P(h)=\delta(h),
\end{equation}
and the limit of stability of this solution is observed by expanding Eq.\ (\ref{eq:cavityeq}) for small $h_i$
\begin{equation}
h_0\simeq \sum_{i=1}^K \tanh(\beta J_{0i})h_i.
\end{equation}
By squaring and taking the average both over $J$ and over $P(h)$ we can see how the size of the distribution (as measured by $\avg{h^2}$, given that $\avg{h}=0$) evolves under the iterations. We get
\begin{equation}
\avg{h^2}'=K\tanh^2(\beta J)\avg{h^2}
\end{equation}
which means that until $T>T_c$ given by (\ref{eq:TcSG}) the stable distribution has a decreasing $\avg{h^2}$ until $\avg{h^2}\to 0$, from which we obtain (\ref{eq:Ppara}) \cite{thouless1986spin}. Proceeding with this reasoning one can also find that for $T\lesssim T_c$:
\begin{equation}
\label{eq:eaop}
q_{EA}\equiv\avg{\avg{\sigma^z_i}^2}_i \propto |T_c-T|.
\end{equation}

Another instructive way to obtain the same results is to consider the spin-spin connected correlation function
\begin{equation}
\label{eq:correlation_function}
C_{ij}\equiv\avg{\sigma^z_i\sigma^z_j}-\avg{\sigma^z_i}\avg{\sigma^z_j}=\frac{1}{\beta}\frac{\partial \avg{\sigma^z_i}}{\partial \eta_j},
\end{equation}
where $\eta_j$ is a local field applied on the spin $j$ (and then sent to zero) $H\to H-\eta_j\sigma^z_j$. By using a telescopic identity and the recursion relations on the field, for $d(i,j)=L\gg 1$ we have 
\begin{equation}
C_{ij}=\beta^{-1}\frac{\partial \avg{\sigma^z_i}}{\partial h_{i-1}}\frac{\partial h_j}{\partial \eta_j}\prod_{k=1}^L\frac{\partial h_{j+k}}{\partial h_{j+k-1}}.
\end{equation}
Now
\begin{equation}
\frac{\partial h_{a}}{\partial h_{a-1}}=\frac{\tanh(\beta J_{a,a-1})\cosh^{-2}(\beta h_{a-1})}{1-\tanh(\beta J_{a,a-1})^2\tanh^2(\beta h_{a-1})}.
\end{equation}
Since we are studying the stability of the paramagnetic phase and the critical region we can set $h_a\to 0$, which gives
\begin{equation}
\frac{\partial h_{a}}{\partial h_{a-1}}\simeq\tanh(\beta J_{a,a-1}).
\end{equation}
Considering that at a distance $L\sim\ln N/\ln K$, diameter of the RRG, there are $K^L$ paths that lead from one spin to another, we have that the susceptibility on a single path has to be summed over all the $K^L$ paths $p$ 
\begin{equation}
C_{ij}=\beta^{-1}\frac{\partial \avg{\sigma^z_i}}{\partial \eta_j}\propto\sum_{p=1}^{K^L}\prod_{a_p=1}^L\tanh(\beta J_{a_p,a_p-1}).
\end{equation}
Since this is a sum of randomly signed, i.i.d.\ terms, we have that the typical value is
\begin{equation}
\sqrt{\avg{C_{i,i+L}^2}}\sim K^{L/2}\tanh(\beta J)^L.
\end{equation}
This decays exponentially if and only if $T>T_c$ in (\ref{eq:TcSG}). Notice that this sum is not dominated by a single term, but it is a collective behavior of the single terms which gives rise to the transition.

This correlation function can be used to define the \emph{shattered susceptibility}
\begin{equation}
\chi_s=\frac{1}{N}\sum_{ij}C_{ij}^2,
\end{equation}
which diverges at the transition and remains infinite in the whole SG phase, below the AT line \cite{de1978stability,thouless1986spin}. On the Bethe lattice this is due to the fact that the exponential growth of the number of sites at distance $r$, $K^r$, dominates over the exponential decay of the typical spin-spin correlation.
 
After mapping this to a directed polymer problem \cite{derrida1988polymers}, one would find that the directed polymer is in the self-averaging phase (see also \cite{ioffe2010disorder}). For comparison, the Anderson model on the Bethe lattice seems always to be in a non-self-averaging (or glassy) phase \cite{pietracaprina16forward}. 

In some sense that will be made clear in the following, for zero temperature the spin-spin correlation can be seen as the propagator of an excitation on the ground state (see Eq. \ref{eq:propagator})
\begin{equation}
C(i,j)=G(i,j|E_0).
\end{equation}
We will see in Section \ref{sect:mean_field} that deep in the paramagnetic phase $\Gamma\to\infty$ we can consider $\sigma^z$ as the operator creating the excitation: $\sigma^z\ket{\rightarrow}=\ket{\leftarrow}$. For finite $\Gamma$ the excitations are dressed but there is still an amplitude of creating an excitation applying $\sigma^z_j$ on $\ket{\Psi_0}$.

\subsection{Computational Complexity}

From a computational perspective, the model (\ref{eq:tfisg}) is interesting because finding or approximating its ground state energy even at $\Gamma = 0$ is a hard optimization problem. For comparison, finding the ground state energy of the same Ising spin glass Hamiltonian defined on planar or toroidal graphs at $\Gamma = 0$ is a problem that can be solved exactly in polynomial time \cite{barahona} (\ie easily) and even when exact solutions are unknown or unlikely to exist one can usually find some approximating algorithm that gives a solution close to the exact one. For example, an $N$-vertex cubic lattice in three dimensions is $O(\epsilon N)$-approximable for any $\epsilon > 0$. In a nutshell, one can partition the lattice into small cubes of size $L \times L \times L$ and define a new Hamiltonian $H'$ by discarding from $H$ the interaction terms that cross from one cube into another. Then $H'$ is a sum of terms defined on different, non-interacting regions of the system and one can solve each term separately, even by brute force (e.g. checking the energy of each possible configuration and taking one with minimal energy). This takes linear time since there are $O(N)$ cubes and each one requires $2^{L^3}$ steps which is a constant in $N$. Finally one gives the ground-state energy of $H'$ as the approximation of the ground-state energy of the original Hamiltonian $H$. The number of bonds discarder by the approximation is given by

\[
\frac{1}{2} \; \lvert \partial_{cube} \rvert \times N_{cubes} = 3L^2 \frac{N}{L^3} = \frac{3N}{L} = O\Big(\frac{N}{L}\Big)
\]

where $N_{cubes}$ is the number of small cubes and $\lvert \partial_{cube} \rvert$ is the number of bonds across the boundary of a single cube. Since $\lvert J_{ij} \rvert = J$ for all bonds $(i,j)$ then the absolute error of the estimated ground-state energy is upper bounded by $3JN/L$ and one can then choose a large enough $L$ so that the $O(N)$ prefactor is as small as desired.

The success of this approximation is based on the fact that surface effects can be neglected since these grow like $L^2$ while volume effects grow like $L^3$. However, this condition is not satisfied in the case of a low-degree regular random graph because RRGs are (on average) good \emph{expander graphs}\cite{Pinsker73}. Roughly speaking, expanders are sparse graphs where, for almost all regions, the boundary of the region (\ie the number of edges connecting the region to the rest of the graph) is proportional to its volume (the number of vertices inside of the region). Physical systems whose interaction graphs are expanders have boundary effects that cannot be neglected. In the cubic lattice example, if we had that $\lvert \partial_{cube} \rvert \propto L^3$ then we would get an absolute error which is independent of $L$ and therefore the previous approximation scheme cannot be fine-tuned to achieve any desired error.

\subsection{Entanglement and Adiabatic Computation}

As a further point of interest, the ground state of (\ref{eq:tfisg}) is expected to be highly entangled somewhere along the $T=0$ line. This is because the topological properties of the interaction structure of a local Hamiltonian such as (\ref{eq:tfisg}) are known to affect the entanglement of its ground state. As very rough way of estimating the entanglement entropy of a region $A$, valid at least for the ground states of gapped Hamiltonians, is to assign a fixed contribution to each interaction that cross from $A$ into the rest of the system. This implies that if many regions of the system are sparsely interacting with the rest of the system (\ie if the spins inside of the region interact with only a few spins on the outside) then the entanglement will be comparatively low on average. On the other hand, if each spin participates in too many interactions then the entanglement in the ground state is suppressed as well, as the monogamy of entanglement prevents spins from being highly entangled with a large number of other spins. A Hamiltonian defined on an expander graph (such as a low-degree regular random graph) seems to lie halfway between these two extremal cases as these graphs are sparse but at the same time almost every subregion is highly-connected with the rest of the graph, and is therefore expected to be highly entangled.

The presence of large entanglement in the ground state makes this model a particularly hard test for the physical implementations of adiabatic quantum computation (AQC), a model of quantum computation where the goal is finding the ground state of a ``problem Hamiltonian'' $H_P$ whose (unknown) ground state encodes the solution to some optimization problem. The computation starts in a quantum system with an unrelated Hamiltonian $H_B$ that has an easy-to-prepare ground state. The Hamiltonian is then changed continuously in time so as to transform $H_B$ into $H_P$ after a finite time. If the adiabaticity condition is satisfied then the system, starting in the ground state of $H_B$, will remain in the istantaneous ground state of the time-dependent Hamiltonian $H(t)$ at all times $t$, and will end up in the ground state of the problem Hamiltonian $H_P$.
AQC has recently attracted attention when the D-Wave company announced the developement of a ``quantum annealer'', a quantum computational device that performs a finite-temperature version of the adiabatic algorithm where the transverse field parameter $\Gamma$ in an transverse-field Ising spin glass Hamiltonian similar to (\ref{eq:tfisg}) starts at $\Gamma \gg J_{ij}$ and is then decreased linearly to zero. While the theoretical description of the adiabatic algorithm assumes that coherence is mantained throughout its entire run, practical implementations such as the D-Wave machine have much shorter decoherence times \cite{zagoskin2014}. This might be problematic, as it was recently shown numerically \cite{bauer2015} that entanglement seems to positively correlate with better performances (\ie better approximations to the ground state energy). Entanglement is also known to play a role in the theory of quantum computational complexity: volume-law entanglement scaling was observed to be associated with the \emph{exponential} speedup that quantum computers are expected to have over classical computers at performing certain specific computational tasks \cite{Orus2004AdiabEntangl}. On the contrary, it is known that efficient quantum algorithms that generate an entanglement that grows slowly with the system size (as measured \eg by the Schmidt rank \cite{Vidal2003EffClassSim} or the maximum number of entangled qubits \cite{jozsa2011}) can achieve at most a polynomial speedup over their classical counterparts.

Since decoherence is one of the main obstacles to building a fault-tolerant quantum computer, we believe that having a quantitative study of the theoretical amount of entanglement that is generated in an adiabatic path will provide a useful way of knowing the degree to which a quantum computer is performing a fully coherent adiabatic quantum algorithm. Consequently, one of the goal of this work is to study the entanglement dynamics of an \emph{ideal} quantum annealing protocol, \ie an actually adiabatic path where the transverse-field coupling constant $\Gamma$ starts at a large value and is then quasi-statically decreased, in order to provide a benchmark for prospective in-depth studies of the entanglement dynamics of a real-life quantum annealer such as the D-Wave machine.

\section{Numerical Study}\label{sect:numerical}

We numerically compute the R\'enyi-2 entropy, quantum Fisher information, Edwards--Anderson order parameter and two-point correlation functions using exact diagonalization for small system sizes $N\approx 20$, and MC simulations for large systems sizes (up to $N=140$). This allows us to obtain information about the thermodynamic-limit properties via finite-size scaling analysis.

\subsection{Methods}
Systems of small size ($ N \le 20 $) are amenable to exact diagonalization (ED) methods, which means that the spectrum of the reduced density matrix is fully accessible.

We use the Lanczos algorithm to extract the ground state of $H$. This step constitutes the bottleneck of the whole procedure, as the Hamiltonian matrix is $ 2^N \times 2^N $ (albeit sparse), which effectively constrains the system size not to exceed $ N \approx 20 $ by too much. The reduced density matrix of half of the system, on the other hand, is only $ 2^{N/2} \times 2^{N/2} $, making it much easier to probe its full spectrum. In this way one can compute the R\'enyi entropy of any desired order $ \alpha $, which in the thermal case is defined as (notice that the von Neumann entropy corresponds to the formal limit $\alpha \rightarrow 1$):
 \begin{equation}
 S_A^{(\alpha)} = \frac{ \log \left[ \Tr \left( \rho_A^\alpha \right ) \right] }{1-\alpha},
 \end{equation}
 where $\rho_A= \Tr_B \rho=\frac{1}{Z}\sum_{s_B}\left<s_A s_B\left|\exp\left(-\beta H\right) \right| s_A s_B\right>$ is the (normalized) reduced density matrix of a subsystem labeled $A$ (an example could be $A=\left\{1,\dotsc,l_A\right\}$, where $l_A$ is the size of the subsystem) obtained by tracing out only the degrees of freedom in the complement subsystem (labeled $B$), and $ \{ \left|s_As_B\right> \}$ is the basis set with the spin degrees of freedom $s_A$ and $s_B$ (corresponding to the complementary regions $A$ and $B$, respectively) separately indicated. The R\'enyi entropy of the ground state is obtained by taking the limit $\beta \rightarrow \infty$.

Another quantity we can probe via exact diagonalization is the ground-state direction-averaged quantum Fisher information (QFI) relative to the total spin operator, namely
\[
	\bar F[\psi_0] \equiv \frac{F[\psi_0; J_x] + F[\psi_0; J_y] + F[\psi_0; J_z]}{3},
\]
where, for $a \in \{x, y, z\}$
\[
	J_a \equiv \frac{1}{2}\sum_{i = 1}^N \sigma_i^a.
\]
For a pure state $\ket{\psi}$ it holds that
\[
	F[\psi; J_a] = 4\big( \bra{\psi}J_aJ_a\ket{\psi} - \bra{\psi}J_a\ket{\psi}^2\big) = 4 \var(J_a)
\]
so knowledge of the ground state $\ket{\psi_0}$ is sufficient for computing $\bar F[\psi_0]$.

\noindent Where exact diagonalization stops being feasible (system sizes $N > 20$) one can compute thermal quantities numerically using the path-integral quantum Monte Carlo (PIMC) approach. In PIMC one uses the path-integral representation of the partition function to map a $D$-dimensional quantum system $H$ to a $(D+1)$-dimensional classical system $H_{\mathrm{eff}}$:

\begin{eqnarray*}
 Z = \Tr e^{-\beta H} &=& \sum_{\vec{s}} \prod_{k=1}^m \braket{s^{(k)}}{e^{-\Delta \beta H}|s^{(k+1)}}\\
 &\approx& \sum_{\vec{s}} e^{-\Delta\beta H_{\mathrm{eff}}(\vec{s})}.
\end{eqnarray*}
Here the set $\{\ket{s^{(k)}}\}_s$ is a basis of the Hilbert space of the system, $\vec{s} = s^{(1)}\dotsb s^{(m)}$, $\Delta\beta \equiv \beta/m$ for some integer $m$ and it is understood that $s^{(m+1)} = s^{(1)}$. The approximation is exact in the limit $m \rightarrow \infty$. The key point is the possibility of rewriting the bra-ket terms as

\[
\braket{s^{(k)}}{e^{-\Delta \beta \hat{H}}|s^{(k+1)}} = e^{-\Delta\beta H^{(k)}(s^{(k)},s^{(k+1)})},
\]
where $H^{(k)}$ is a \emph{classical} energy term. Then $H_{\mathrm{eff}} \equiv \sum_k H^{(k)}$. Expected values of quantum observables $\mathcal{O}$ can then be computed in a straightforward manner by applying this quantum-to-classical map to the expression 

\[
\langle \mathcal{O} \rangle = \Tr(\mathcal{O} e^{-\beta H})/Z,
\] 
and then using standard Monte Carlo sampling on the resulting classical system. Ground-state properties are accessible in the limit $\beta \rightarrow \infty$. We use this standard PIMC approach when computing the Edwards--Anderson order parameter (\ref{eq:eaop}) and the spatial correlation functions (\ref{eq:correlation_function}).

However, one must take a somewhat different approach when computing entanglement entropies.  R\'enyi entropies are \emph{not} observables of the system (trivially, they are not linear operators) hence a naive PIMC is inapplicable. In order to compute the R\'enyi-$2$ entanglement entropy of a region $A$ of the system we use the replica-approach algorithm described in \cite{roscilde} that allows for computations of nonlinear quantities.
 
When computing the R\'enyi-2 entropy using the Monte Carlo approach described before we set an inverse temperature of $\beta = 15$ in order to project the thermal state to the ground state and we use $m=150$ timeslices for each replica so that the imaginary time discretization is given by $\Delta \beta = 1/10$. We also add a weak longitudinal field term $-h\sum_i \sigma_i^z$ with $h = 0.05$ to the Hamiltonian (\ref{eq:tfisg}) since we noticed that it shortens the Monte Carlo convergence times. The (small) effect that this field has on the R\'enyi entropy values is expained in Subsection \ref{subsect:results}. In order to assign an error bar to the result for each disorder realization we performed ten independent simulations starting from a different initial configuration (the same for all replicas). In this case, the value of $S^{(2)}_A$ for each realization of disorder is obtained by averaging over the ten results. For the Edwards--Anderson order parameter and the correlation functions we use $\beta = 40$ and $\Delta \beta = 1/8$.

We emphasize that our analysis focuses on the regime of large and intermediate values of $\Gamma \gtrsim \sqrt{K} > 1$ where $K$ is the connectivity of the RRG. Here, for the finite system size we consider, the problems due to the slowdown of the Monte Carlo dynamics in the glassy phase are not overwhelming. More details on all these techniques are contained in the appendix.

\subsection{Average over the disorder}

Our system contains disorder and our numerical methods only work on a fixed realization of disorder at a time. The natural way of taking a disorder average is then to generate a large number of realizations, compute the desidered quantity for each realization and then take the average of the results. When computing these average values for a fixed system size $N$ we have to consider two distinct disorder variables: \emph{i)} the topological structure of the interaction graph $\mathcal{G} = (V,E)$ and \emph{ii)} the couplings $J_{ij}$ in the interaction Hamiltonian. A specific instance of the simulation is defined by the values assigned to these two variables. To create such an instance $i_N = i_N(\mathcal{G},J_{ij})$ for a system of $N$ spins we generate an $N$-vertex regular random graph using the Steger and Wormald algorithm \cite{steger1999generate} and then we randomly assign a coupling $J_{ij} = \pm 1$ to each edge of the graph, with equal probability.
  
 In the case of the R\'enyi-2 entropy an instance is defined by \emph{three} variables $i_N = i_N(\mathcal{G},J_{ij},A)$, since one has to take into account also the region $A$ of which we compute the R\'enyi entropy. This region is generated according to the Random Region protocol described in Fig. (\ref{fig:randomregion}).

\subsection{Results}\ref{subsect:results}
\emph{R\'enyi entropy ---} Our first goal is understanding the thermodynamic-limit behaviour of the disorder-averaged R\'enyi-2 entropy $\overline{S}_A^{(2)}$ on the $T=0$ line of the $(T,\Gamma)$-phase diagram of the Hamiltonian (\ref{eq:tfisg}). We do this by computing the finite-size values $\overline{S}_A^{(2)}(N)$ and then studying the limit $N \rightarrow \infty$. 

\begin{figure}
  \includegraphics[width=\linewidth]{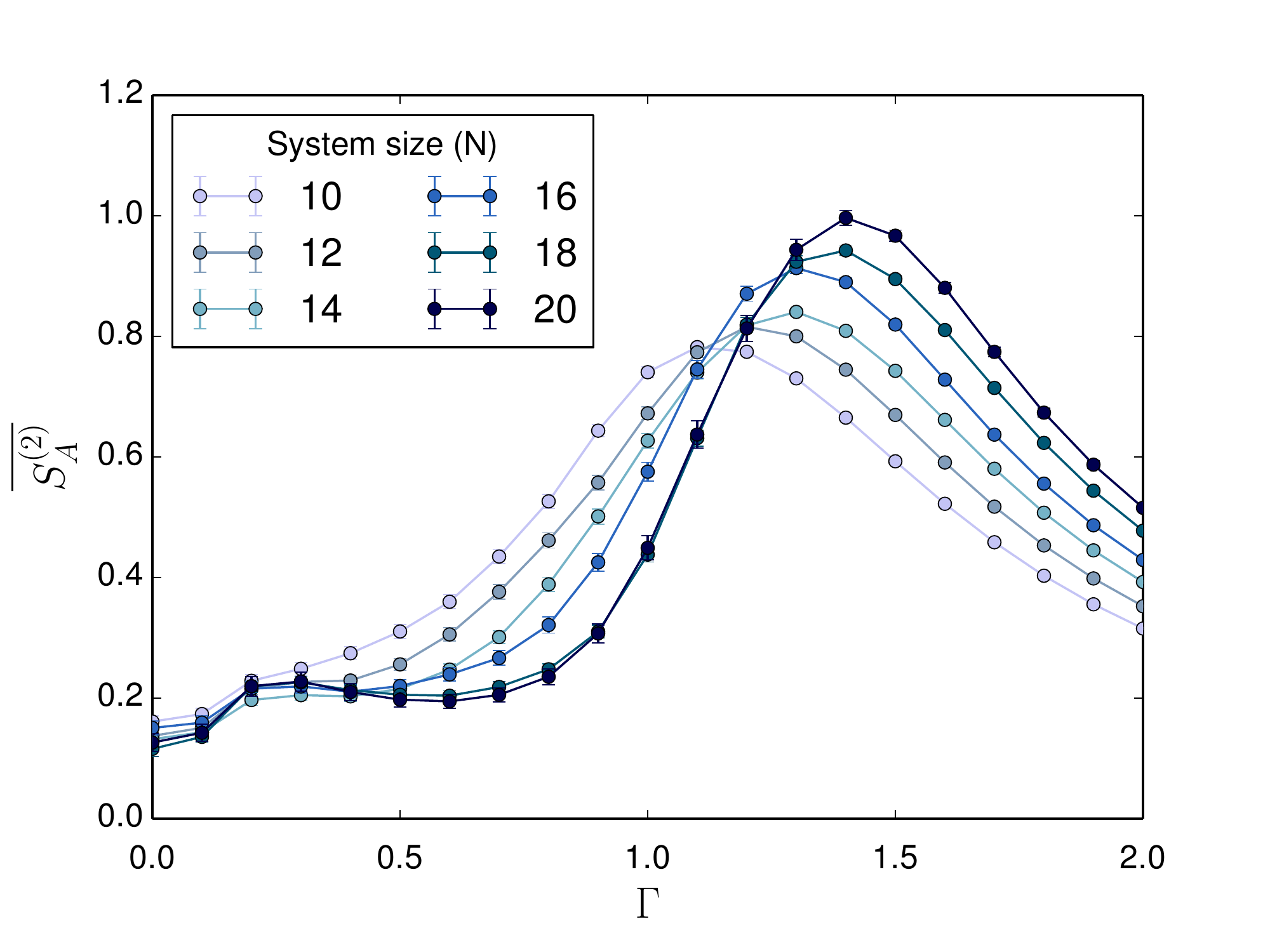}
  \caption{Plot of the average value of the R\'enyi entropy $\overline{S_A^{(2)}}$ as a function of the transverse field strength $\Gamma$, for different sizes of the system (exact diagonalization results).}
  \label{fig:ed_renyi}
\end{figure}

\begin{figure}
  \includegraphics[width=\linewidth]{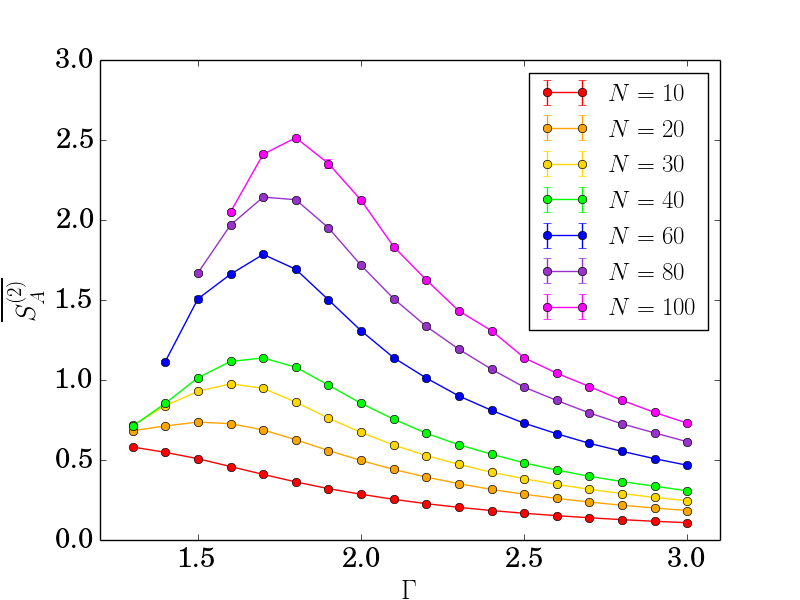}
  \caption{Plot of the average value of the R\'enyi entropy $\overline{S_A^{(2)}}$ as a function of the transverse field strength $\Gamma$, for different sizes of the system (quantum Monte Carlo results).}
  \label{fig:renyi}
\end{figure}

Small-sizes results obtained by exact diagonalization show a volume-law scaling for all values of $\Gamma$ (Fig. \ref{fig:ed_renyi}), which is confirmed by the Monte Carlo results (Fig. \ref{fig:renyi}). Comparing the QMC and the ED results we note that the weak longitudinal field we use in the Monte Carlo simulations reduces the numerical value of the R\'enyi entropy in the critical region while leaving the positions of the peaks essentially unaffected. Moreover, we verified that ED and QMC results agree on system sizes where both can be applied when the weak longitudinal field is included also in the ED calculations. We see that, for each system size $N$, the $S_A^{(2)}$ curve attains a peak. These peaks of the entanglement entropy have been associated to criticality in quantum phase transitions \cite{Fazio2008review}, where the position of the maximum in the thermodynamic limit is the critical point of the transition. In order to extract this value from our finite-size data we define, for each finite $N$, the estimator $\Gamma_c(N)$ as the point where the $S_A^{(2)}(N)$ curve attains its maximum and then we study the limit as $N \rightarrow \infty$.

\begin{figure}
	\begin{framed}
		\textbf{RANDOM REGION}
		
		\vspace{8pt}
		
		We select a connected region $A$ of $N/2$ vertices in the following way

		\begin{itemize}
			\item[1)] Select a vertex $v$ uniformly at random and define $A \equiv \{v\}$.
			\item[2)] Given the region $A$, select uniformly at random a vertex $u$ from the boundary $\partial A$ of the region $A$. Then update the region $A \mapsto A' \equiv A \cup \{u\}$.
			\item[3)] Repeat point 2) until the desired size of $A$ is reached.
		\end{itemize}
	\end{framed}
	\caption{Protocol for the generation of a random region $A$.}
  \label{fig:randomregion}
\end{figure}

\noindent We approximate the peaks by fitting the numerical data with parabolic curves and then we take the the vertices of the parabolas to be the finite-size estimators $\Gamma_c(N)$. Then we use a simple $1/N$ fitting ansatz to extract $\Gamma_c \equiv \lim_{N \rightarrow \infty}\Gamma_c(N)$ by looking at the convergence of these peaks in the thermodynamic limit 

\begin{equation}
\Gamma_c(N) \equiv \Gamma_c + \frac{\Delta\Gamma_c}{N}.
\end{equation}

\begin{figure}
  \includegraphics[width=\linewidth]{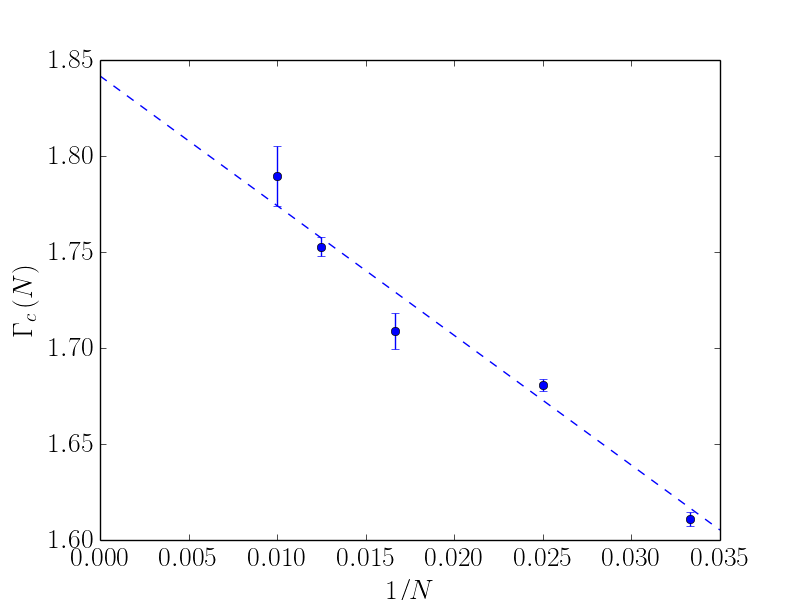}
  \caption{Estimated maxima of $S_A^{(2)}$ are fitted with a $1/N$ scaling behaviour.}
  \label{fig:gamma_c_renyi}
\end{figure}

\begin{figure}
  \includegraphics[width=\linewidth]{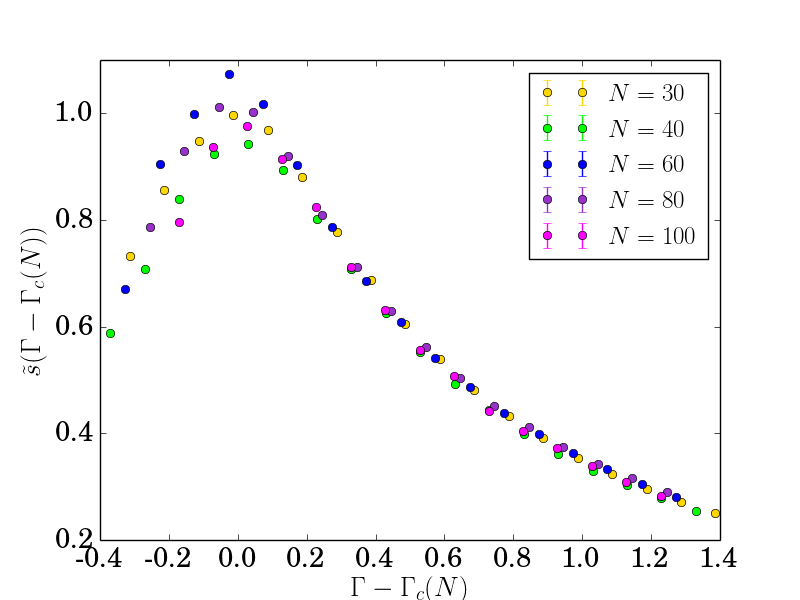}
  \caption{Data collapse for the finite-size scaling. Finite-size estimates to the critical point $\Gamma_c$ are given by the fit $\Gamma_c(N) = \Gamma_c + \Delta\Gamma_c/N$.}
  \label{fig:fss}
\end{figure}

\noindent We obtain a critical point of $\Gamma_c = 1.84 \pm 0.02$ (Fig. \ref{fig:gamma_c_renyi}). By considering both the critical scaling of $S_A^{(2)}$ and the finite-size shift on $\Gamma_c(N)$ we propose the finite-size scaling (FSS) ansatz

\begin{equation}
S^{(2)}(N,\Gamma) = \tilde{s} \Big(\Gamma - \Gamma_c(N)\Big)(aN + b),
\end{equation}

which gives the data collapse shown in Fig. (\ref{fig:fss}). We consider $S_A^{(2)}$ data only in the regime where the Edwards--Anderson order parameter (see discussion below) is small.

\vspace{10pt}

\begin{figure} 
	\centering
\includegraphics[width=\linewidth]{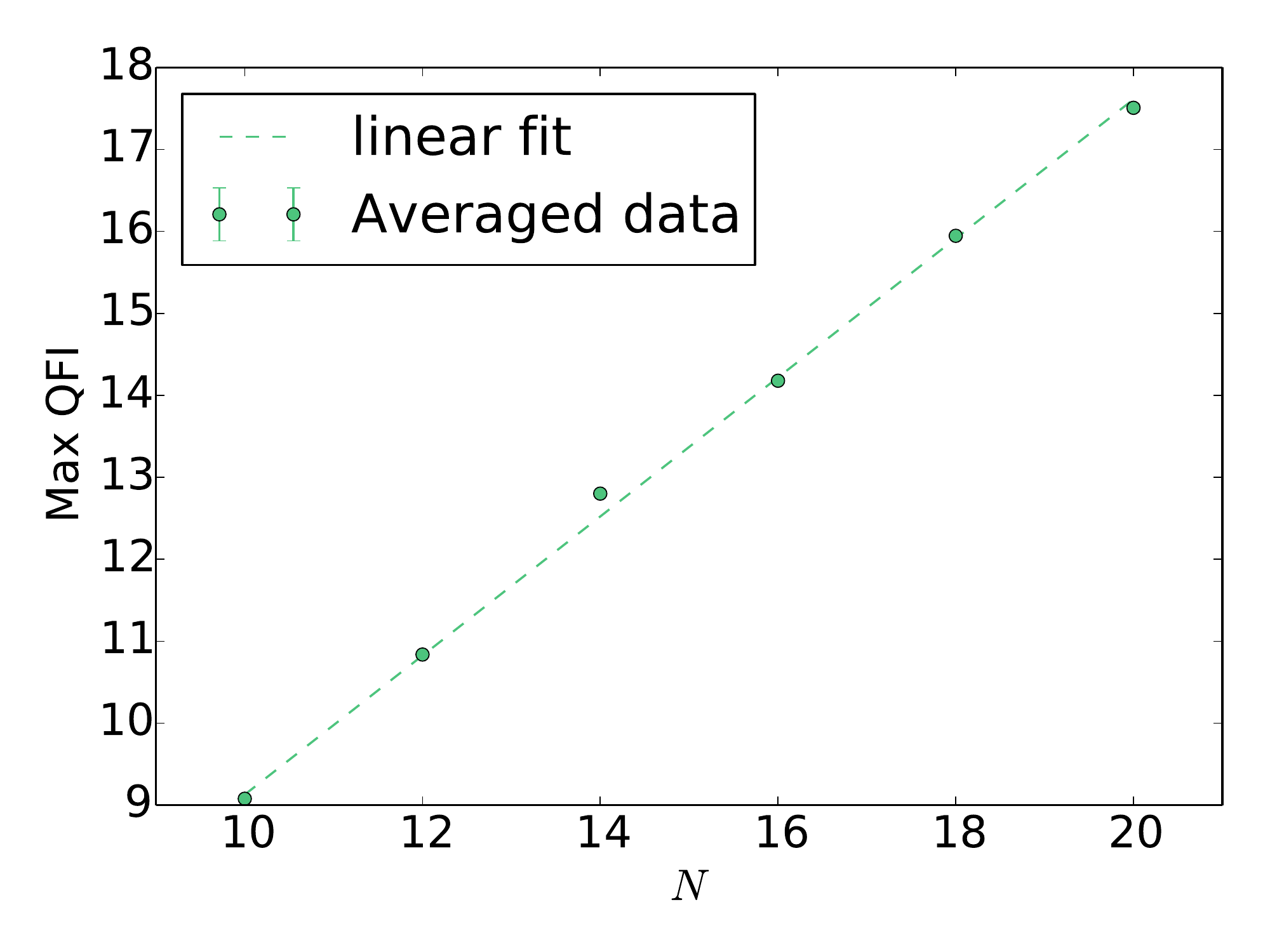}
	\caption{Linear fit of the ground state QFI peak.}
	\label{fig:aqfi_y}
\end{figure}

\emph{Quantum Fisher Information ---} We compute the Quantum Fisher Information (QFI) by exact diagonalization for system sizes $N \leq 20$. The QFI curves have a shape reminiscent of the entanglement entropy curves, including the peak at \( \Gamma_c(N) \). A FSS analysis of the peak leads to Fig. \ref{fig:aqfi_y}, confirming a linear size scaling with linear coefficient \( a = \SI{0.85(1)}{} \).

\begin{figure}
	\centering
	\includegraphics[width=\linewidth]{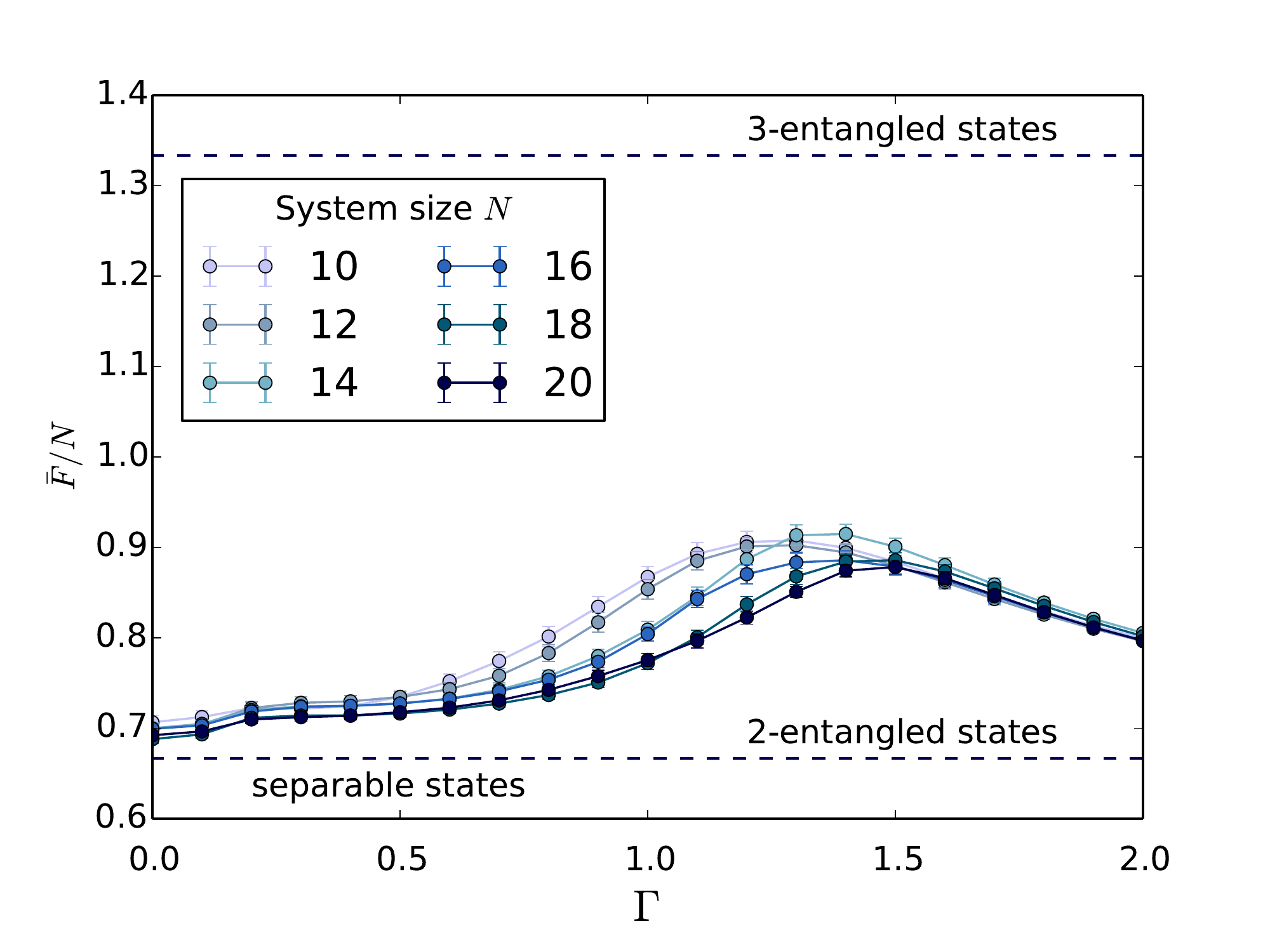}
	\caption{Disorder-averaged ground-state quantum Fisher information density \( \bar F/N \) as a function of the transverse field strength $\Gamma$. The lower dashed line represents the separability criterion violation: states above it are entangled. States above the upper dashed line are at least 3-entangled.}
	\label{fig:aqfi_comparison}
\end{figure}

 It can be shown that the condition
 \begin{equation}\label{f-criterion}
 	\bar F[\psi] > \frac{1}{3}\left[s(k^2 + 2 k - \delta_{k,1}) + r^2 + 2 r - \delta_{r,1}\right],
 \end{equation}
 where $ s = \lfloor N/k \rfloor $ and $ r = N - s k $, implies that the state $ \psi $ contains multipartite entanglement between at least $k+1$ spins \cite{hyllus2012}, \ie it cannot be written as $\ket{\psi} = \bigotimes_i \ket{\psi_i}$ where each $\ket{\psi_i}$ is a state of $n_i \leq k$ spins. 

 Note that for \( N \) even, proving 3-entanglement from Eq. \eqref{f-criterion} requires \( \bar F[\psi_0]/N \ge 4/3 \), which FSS suggests will never be satisfied (since \( 0.85 < 4/3 \), see\ Fig. \ref{fig:aqfi_comparison} and \ref{fig:aqfi_y}): the entanglement in the ground state seems to be limited to two-spins states. We emphasize that the QFI $F[\cdot, \hat{O}]$ is dependent on the specific choice of the generator $\hat{O}$ so, strictly speaking, the previous results are to be taken as \emph{lower bounds} to the multipartite entanglement present in the ground state. The actual value is obtained by maximizing $F[\cdot,\hat{O}]$ over all possible choices of $\hat{O}$. We are unable to solve this maximization problem but our choice of generator seems to be a natural one when considering only one-spin operators.

\vspace{10pt}

\emph{Edwards--Anderson order parameter ---} In order to characterize the glassy properties on the low $\Gamma$ phase, and also to cross-check our estimate for the critical point $\Gamma_c$, we compute the quantity
\begin{equation}\label{eq:QEAOP}
q_{EA} \equiv \frac{1}{N} \sum_{i=1}^N \langle \sigma_i^z \rangle^2,
\end{equation}
in the ground state of the Hamiltonian (\ref{eq:tfisg}) over a $3$-regular random graph. This is the Edwards--Anderson order parameter, introduced in \cite{edwardsanderson1975} as the order parameter for a glassy phase transition: in systems with such a transition it holds that $q_{EA} = 0$ in the paramagnetic phase while $q_{EA} > 0$ in the glassy phase. For each specific realization of the disorder $i_N = i_N(\mathcal{G},J_{ij})$ we obtain (\ref{eq:QEAOP}) by computing each value of $\langle \sigma_i^z \rangle$ via Monte Carlo simulations. We then take the average over the disorder obtaining a curve $q_{EA}^{(N)}(\Gamma) \equiv q_{EA}(\Gamma,N)$ at fixed size $N$ (results are shown in Fig. \ref{fig:qea_pure}). The critical point $\Gamma = \Gamma_c$ is then the point of singularity of the curve 

\[
q_{EA}(\Gamma) = \lim_{N\rightarrow \infty} q_{EA}^{(N)}(\Gamma).
\]

\begin{figure}
  \includegraphics[width=\linewidth]{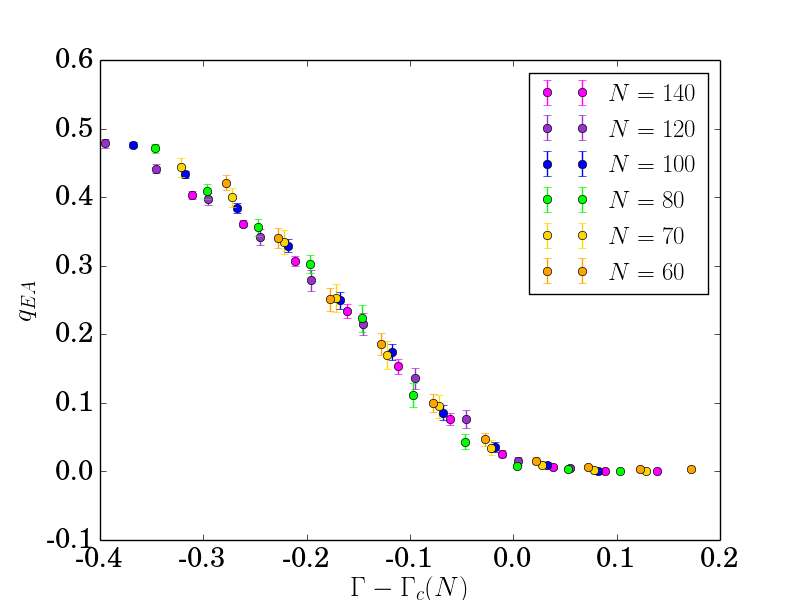}
  \caption{Data collapse of the curves of Fig. (\protect\ref{fig:qea_pure}). The scaling for $\Gamma_c(N)$ is shown in Fig. \ref{fig:critical_point}.}
  \label{fig:qea_fss}
\end{figure}

We observe strong finite-size effects and most curves have smooth transitions from a region where they are zero (within statistical error) to a region where they attain positive values. Therefore, for any fixed size $N$, our finite-size estimates of the critical point $\Gamma_c(N)$ are obtained in the following way. First we shift all of the curves horizontally so that they fall on top of each other, as in Fig. (\ref{fig:qea_fss}). Then we take a linear fit of the accumulated data around the point where $q_{EA}$ starts being finite and obtain a slope $s$. Finally we take linear approximations to each of the $q_{EA}^{(N)}$ curves in Fig. (\ref{fig:qea_pure}) using the fixed slope $s$. The values $\Gamma_c(N)$ are defined as the $x$-intercept of these new linear fits. The ansatz

\[
\Gamma_c(N) = \Gamma_c + \frac{\Delta\Gamma_c}{N}
\]
for these points gives $\Gamma_c = 1.82 \pm 0.02$. We get an excellent agreement with the critical point we extracted from the study of the R\'enyi entropy. We see this as an indication that the QPT we detected using the R\'enyi entropy is exactly the glassy phase transition of the model.

\begin{figure}
  \includegraphics[width=\linewidth]{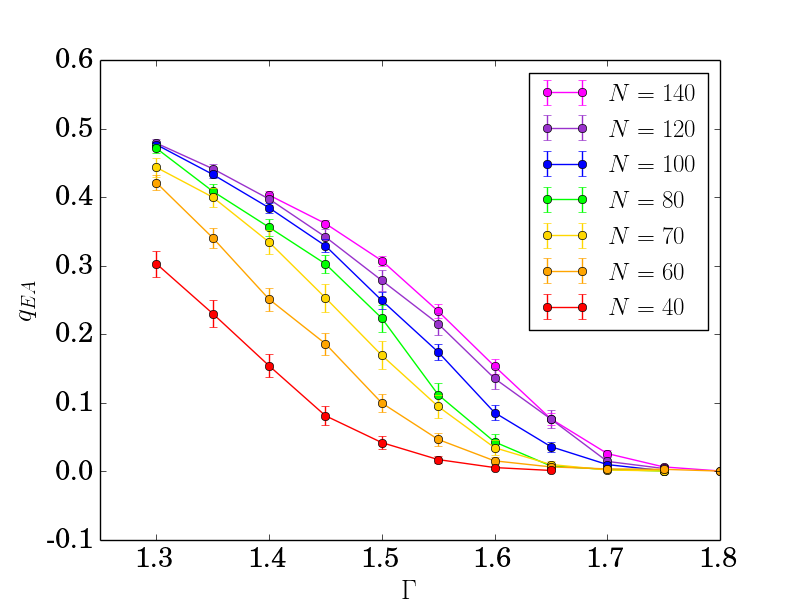}
  \caption{Disorder-averaged Edwards--Anderson order parameter $q_{EA}$ as a function of the transverse field strength $\Gamma$, for different system sizes. The finite-size effects induce a smoothing-out of the curves that disappears as $N \rightarrow \infty$.}
  \label{fig:qea_pure}
\end{figure}

\vspace{10pt}

\begin{figure}
  \centering
	    \includegraphics[width=0.9\columnwidth]{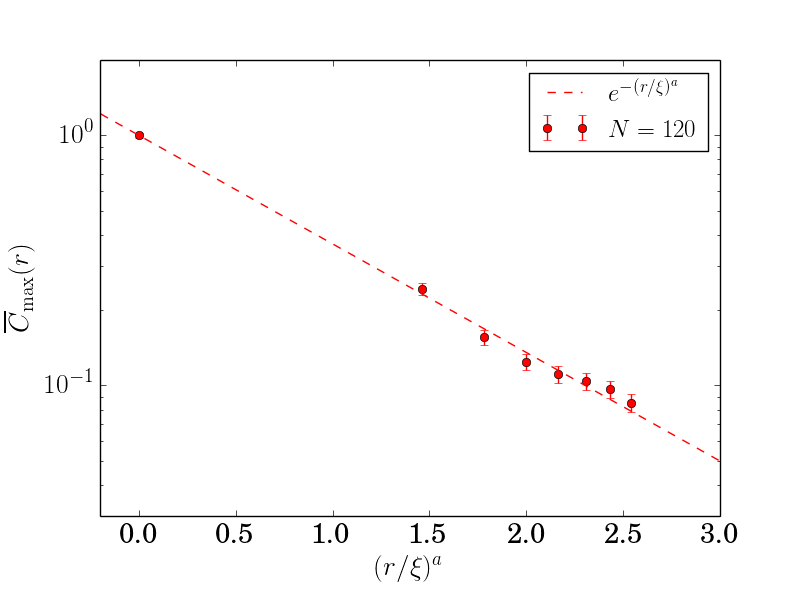}
  \caption{Stretched expenential fit for the disorder-averaged maximal correlation $\overline{C}_{\max}$ for system size $N = 120$ and $\Gamma = 1.50$. Fitting parameters are $a = 0.28$, $\xi = 0.25$. Other values of $N$ and $\Gamma$ give qualitatively similar results, with different values for the fitting parameters $a,\xi$.}
  \label{fig:stretched_exponential}
\end{figure}

\begin{figure}
	\centering
	\includegraphics[width=0.9\columnwidth]{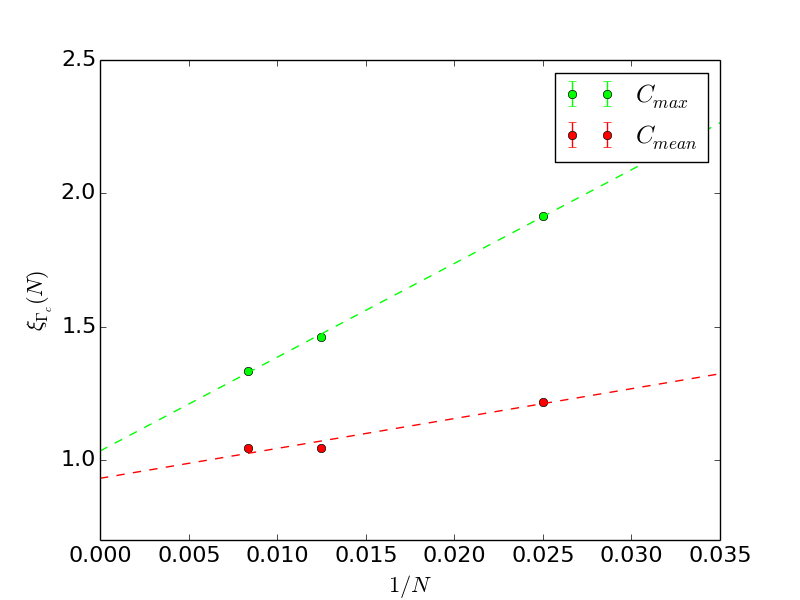}
  \caption{$1/N$ fit of the critical correlation lengths obtained from $C_{\max}$ and $C_{\mean}$.}
	  \label{fig:xi_critical}
\end{figure}

\emph{Connected correlations ---} We compute the connected correlation function
\[
 C_{ij} \equiv \langle \sigma_i^z \sigma_j^z \rangle - \langle \sigma_i^z \rangle \langle \sigma_j^z \rangle
\]
of the Hamiltonian (\ref{eq:tfisg}) as follows: for each realization of disorder we randomly choose a central spin $i$ on the interaction graph and we compute all connected correlation functions $C_{ij}$ for all sites $j$ in the system. Then for any integer $r$ we define the mean correlation at distance $r$, $C_{\mathrm{mean}}(r)$, and the maximal correlation at distance $r$, $C_{\max}(r)$, by taking respectively the average and the maximum of (the absolute value of) the correlations among all sites $j$ that are $r$ steps away from the central spin $i$ in the distance of the graph $\mathcal{G} = (V,E)$

\begin{eqnarray*}
C_{\mathrm{mean}}(r) &\equiv& \mean \{ \: \lvert C_{ij}\rvert : j \in V, \operatorname{dist}(i, j) = r \}, \\	
C_{\max}(r) &\equiv& \max \{ \: \lvert C_{ij}\rvert : j \in V, \operatorname{dist}(i, j) = r \},
\end{eqnarray*}
then we average these quantities over many realizations of disorder to get the average mean correlation $\overline{C}_{\mathrm{mean}}(r)$ and the average maximal correlation $\overline{C}_{\max}(r)$ as functions of the distance $r$. We find that both $\overline{C}_{\mathrm{mean}}$ and $\overline{C}_{\max}$ follow a stretched-exponential behaviour
\[
\overline{C}_{\mathrm{mean},\mathrm{max}} \sim e^{-(r/\xi)^a}
\]
at any transverse field value $\Gamma$ (see Figure \ref{fig:stretched_exponential}). Stretched exponentials are usually encountered in disordered systems when taking the disorder average of an exponentially-decaying quantity $Q$ that has different exponential decay constants $\xi_i$ for each realization of disorder $i$, \ie $Q_i(r) \sim e^{-r/\xi_i}$. Therefore we argue that our model shows a distribution of correlation lengths $\xi_i$ between different disorder realizations. By defining $\Gamma_c(N)$ as the value of $\Gamma$ where the curve $\xi(\Gamma)$ for size $N$ attains its maximum, we note that for both quantities $\overline{C}_{\mathrm{mean}}$ and $\overline{C}_{\mathrm{max}}$, the critical correlation lengths $\xi_{\Gamma_c}(N)$ we obtained from the fits are \emph{decreasing} with the system size $N$, and converge to a finite value in the thermodynamic limit (Fig. \ref{fig:xi_critical}).

\begin{figure}
	    \includegraphics[width=0.9\columnwidth]{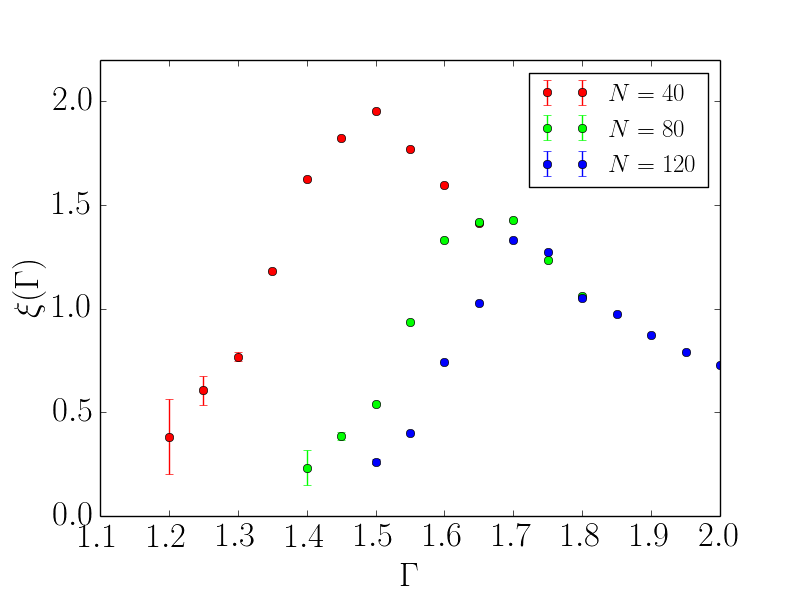}
  \caption{Correlation lengths $\xi(\Gamma)$ obtained from the stretched-exponential fit to the numerical results of $\overline{C}_{\mathrm{max}}$. Finite-size estimators $\Gamma_c(N)$ for the critical point are defined as the vertices of parabolic fits to the curves.}
	  \label{fig:xi}
\end{figure}

\vspace{10pt}

\begin{figure}
	\centering
	\includegraphics[width=0.9\columnwidth]{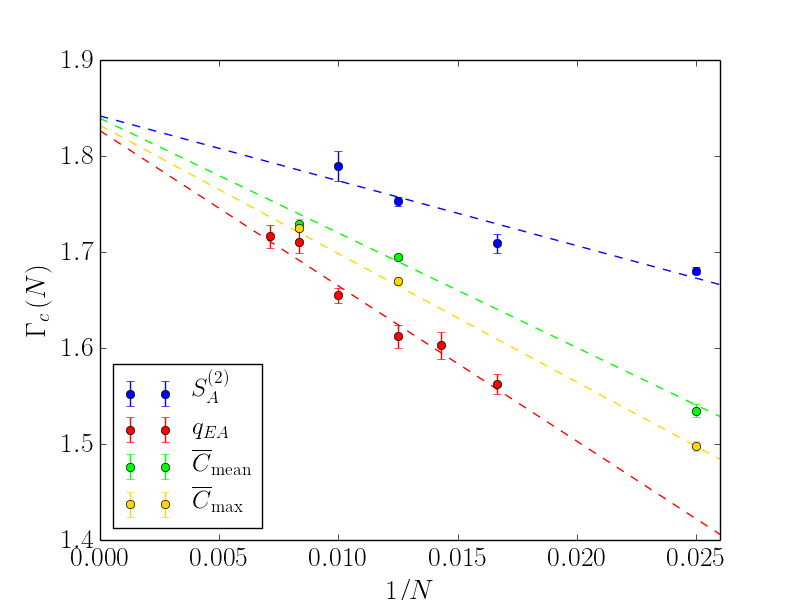}
  \caption{$1/N$ fits of the finite-size estimators $\Gamma_c(N)$ for the critical point obtained from all the quantities considered in this work. While finite-size corrections differ between quantities, all seem to indicate a critical point $1.82 \leq \Gamma_c \leq 1.85$.}
	  \label{fig:critical_point}
\end{figure}

\emph{Critical Point ---} Summarizing our results, we have computed four independent ways of estimating the critical point of the glassy phase transition, \ie four sets of finite-size estimators $\Gamma_c(N)$. Each set was extracted from the study of a different physical quantity whose behaviour is known to be affected by criticality. Finite-size corrections disappear as $1/N$ and in the thermodynamical limit all estimates seem to converge to a critical point $1.82 \leq \Gamma_c \leq 1.85$, as shown in Fig. (\ref{fig:critical_point}).

\section{A mean field model of the transition}\label{sect:mean_field}

The fact that $\Gamma_c/J\gtrsim 1$ (and, as we will see, that it grows with $K$), should suggest us that a better starting point for analyzing the transition is $\Gamma=\infty$ rather than $\Gamma=0$, which has been the method adopted in previous investigations of the glassy phase. Assuming, as we will argue, that if a MBL phase exists for this model then it must be located inside of the glassy phase, an expansion around $\Gamma=0$ is more likely to break down at the MBL transition rather than the spin glass transition.

We are going to show that a simple mean field theory, obtained by dressing the excitations around $\Gamma=\infty$, gives a transition point quite close to the unbiased prediction obtained from the Monte Carlo simultions and predicts two main things: a) the scaling of the critical point $\Gamma_c\propto\sqrt{K}$ and b) the exponential decay of correlations inside the paramagnetic phase.

First of all notice that the spectrum for $J=0$ is composed of $N$ bands labeled from $n=0$ to $n=N$ depending on the number of spins oriented in the $-x$ direction since $\sigma_x\ket{\rightarrow}=\ket{\rightarrow}$ and $\sigma_x\ket{\leftarrow}=-\ket{\leftarrow}$. The ground state is the only state at $n=0$ and it is
\begin{equation}
\ket{0}\equiv\ket{\rightarrow,\rightarrow,\dotsc,\rightarrow},
\end{equation}
and we take it to have energy $E=0$.

The band of the first excited state has energy $2\Gamma$ and can be labeled as
\begin{equation}
\ket{i}\equiv\ket{\rightarrow,\dotsc,\leftarrow,\dotsc,\rightarrow}
\end{equation}
where $i$ is the position of the $\leftarrow$ spin. And so on we have $\ket{i,j}$, $\ket{i,j,k}$ etc.\ at energy $4\Gamma, 6\Gamma, \dotsc$ etc.

\begin{figure}[htbp]
\begin{center}
\includegraphics[width=\linewidth]{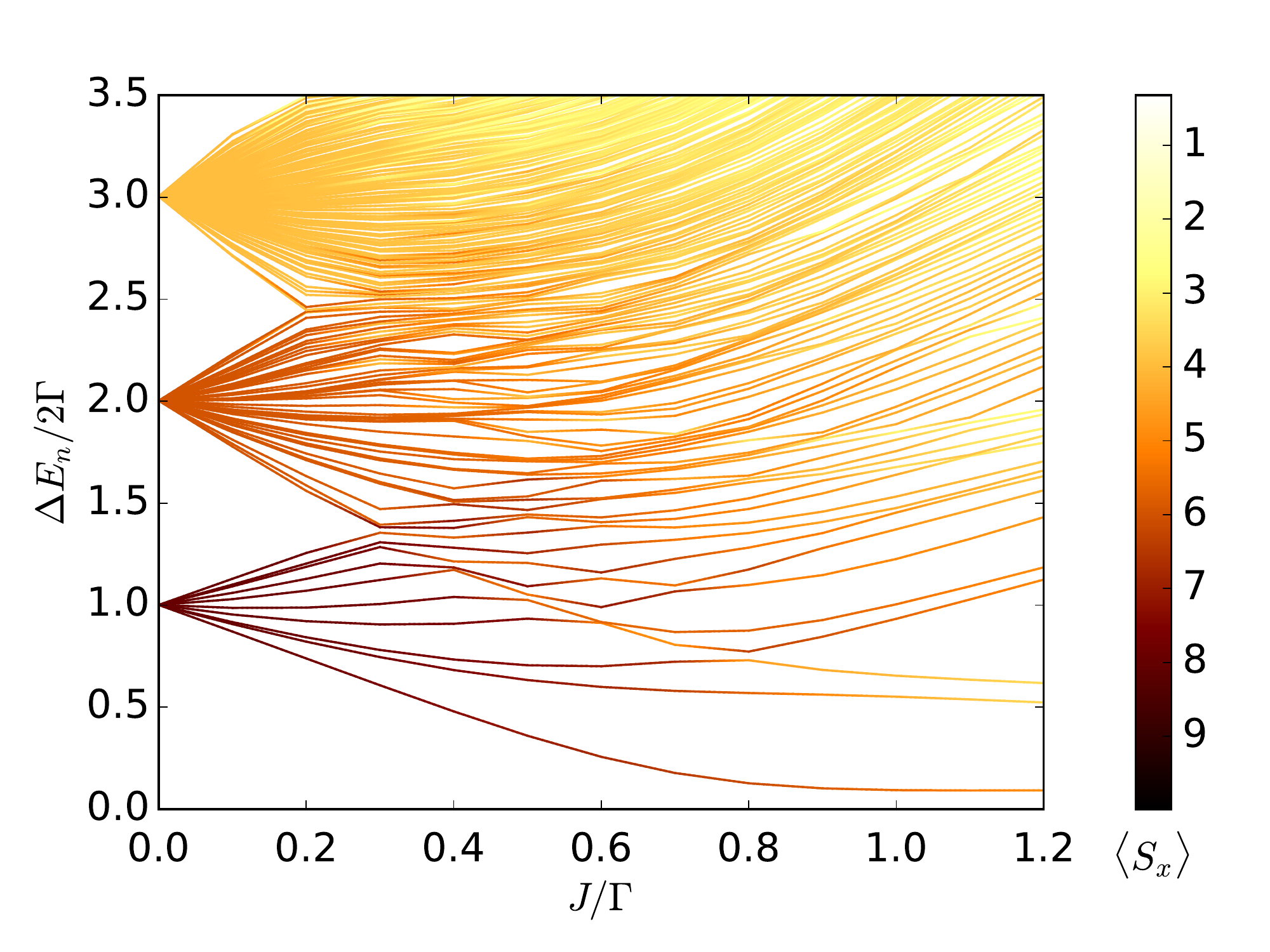}
\caption{Perturbative lifting of the degeneracy of the bands in the spectrum of the transverse-field Hamiltonian $H_0$ (at fixed $\Gamma$) as $J$ is increased from zero. Color code denotes the average value of $S_x \equiv \sum_{i} \sigma_i^x$ of the state associated with the energy level.}
\label{fig:default}
\end{center}
\end{figure}

The operator $\sigma^z_i\sigma^z_j$ flips two $x$ spins so it can do three things: when it hits two $\rightarrow$ states it changes the sector by $\Delta n=2$, therefore changing the energy by $\Delta E=4\Gamma$; when it hits one $\rightarrow$ and one $\leftarrow$, it moves the $\leftarrow$ from $i$ to $j$ or from $j$ to $i$ with amplitude $J_{ij}$; when it hits two $\leftarrow$'s it annihilates them thereby changing the energy by $\Delta E=-4\Gamma$. 

Of these processes, for $\Gamma\gg J$ only the second one has to be considered to lowest order. Therefore we have a set of $n$ particles which have hard-core interactions. It is instructive to start with the single particle sector which describes the low-energy excitations around $\Gamma=\infty$. The single particle sector has a particle hopping on the interaction graph with Hamiltonian
\begin{equation}
H^{(1)}=-\sum_{ij}J_{ij}\ket{i}\bra{j}+\textrm{h.c.}\ .
\end{equation}
Such a random Hamiltonian on the Bethe lattice can be studied by a cavity recursion relation which depends only on $J_{ij}^2$ so we can see straight away that the spectrum has to be the same as the that of the adjacency matrix of the RRG with hopping $J$ and, in particular, the ground state energy is a self-averaging quantity:
\begin{equation}
E^{(1)}_0=-2J\sqrt{K}.
\end{equation}
It is known that all the states, including the ground state, of this disordered Hamiltonian are delocalized. For more details see \cite{PhysRevE.90.052109}.

We see how by increasing $J$ this state gets closer to the ground state. If we neglect processes of $O(J^2/\Gamma^2)$ which dress both the ground state and the we get a crossing when $E^{(1)}+2\Gamma=0$, or
\begin{equation}
\label{eq:gammac1}
\Gamma^{(1)}_c=J\sqrt{K}\simeq 1.41 J.
\end{equation}

This is an underestimate, but quite a good one, of the critical point as observed in the numerics $1.82 \leq \Gamma_c \leq 1.85$. 
Notice also that this scaling of $\Gamma_c$ with $K$ is important as the limit $K\to\infty$ should reproduce the fully connected, Sherrington-Kirkpatrick model \cite{bray1980replica,goldschmidt1990ising,PhysRevB.39.11828}, by scaling $J\to J/\sqrt{K}$. This scaling keeps the transition temperature $T$ finite and we find here that the independent particle picture also returns a finite $\Gamma$.

We now try to go beyond the independent particle approximation. By going to higher $n$ sectors we need to solve the problem with $n$ particles which, to first order in $J/\Gamma$, are hard-core bosons. This is a complicated problem (which, by the way, is equivalent to replacing $\sigma_i^z\sigma^z_j\to\sigma^+_i\sigma^-_j$). If $n\ll N$, as a first approximation we can assume the particles as non-interacting. This gives for the ground state of the $n$-th sector $E^{(n)}_0=-2nJ\sqrt{K}$, and by matching $E^{(n)}+2n\Gamma=0$ we find the very same result (\ref{eq:gammac1}). 

In the spirit of mean-field theory we can add the interaction of the particles by observing that, when considering only two-particles processes, the process $J\sigma^z_i\sigma^z_j\ket{\leftarrow,\leftarrow}=\ket{\rightarrow,\rightarrow}$ which reduces by $4\Gamma$ the energy of the state, can be interpreted as a zero-range attraction potential energy $V=-4\Gamma$ when two particles are on top of each other. This correction to the energy of the ground state goes in the right direction, increasing the value of $\Gamma_c$. In fact, we have an expected energy per particle $V\rho/2=-2\Gamma \rho$ where $\rho=n/N\ll 1$ is the probability to find two particles on the same site, assuming the particles are indeed delocalized
\begin{equation}
E^{(n)}/n=-2J\sqrt{K}-2\Gamma \rho,
\end{equation}
and therefore the equation defining the critical point is $-2J\sqrt{K}-2\Gamma \rho+2\Gamma=0$
so
\begin{equation}
\Gamma_c^{mf}=J\sqrt{K}(1+\rho-O(\rho^2)).
\end{equation}
The modification is in the right direction: increasing the density $\rho=n/N$, the value of $\Gamma_c$ at the critical point increases; moreover the $\sqrt{K}$ scaling is preserved. Higher and higher $n$ sectors cross the ground state at larger $\Gamma$. Of course this approximation is reliable only if $\rho \ll 1$ (agreement with the observed $1.82 \leq \Gamma_c \leq 1.85$ requires $\rho\simeq 0.3$, close to the center of the band $\rho=1/2$). 

Let us now see how to connect the computed spin--spin correlation function with the propagator of the particle on the Bethe lattice.

From the definition
\begin{equation}
C(i,j)=\bra{\Psi_0}\sigma^z_i\sigma^z_j\ket{\Psi_0}-\bra{\Psi_0}\sigma^z_i\ket{\Psi_0}\bra{\Psi_0}\sigma^z_j\ket{\Psi_0}
\end{equation}
we see that we can write the correlation function as the susceptibilty of the magnetization at $i$ with respect to a magnetic field $\eta_j$ at $j$:
\begin{equation}
C(i,j)=\frac{\partial}{\partial \eta_j}\bra{\Psi_0(\eta_j)}\sigma^z_i\ket{\Psi_0(\eta_j)}\Big|_{\eta_j\to 0},
\end{equation}
where $\Psi_0(\eta_j)$ is the ground state with the field at $\eta_j$. By perturbation theory this is
\begin{equation}
\ket{\Psi_0(\eta_j)}=\ket{\Psi_0}+\eta_j\sum_{n>0}\ket{\Psi_n}\frac{\bra{\Psi_n}\sigma^z_j\ket{\Psi_0}}{E_n-E_0}+O(\eta_j^2)
\end{equation}
and after rewriting the denominators as integrals over time we define $\sigma_j^z\ket{\Psi_0}=\ket{\tilde{j}}$ is the position ket of an excitation at $j$. Notice that, to lowest order in $J/\Gamma$, $\ket{\tilde{j}}$ is exactly the single $x$-spin flipped state $\ket{j}$ we introduced before. So, ignoring this difference, we can write
\begin{equation}
C(i,j)=i\int_0^\infty dt\ e^{iE_0t}\bra{i} e^{-iHt}\ket{j},
\end{equation}
\ie the spin--spin correlation function is the frequency component $E_0$ of the propagator of a particle created at $j$ and detected at $i$:  
\begin{equation}\label{eq:propagator}
C(i,j)=G(i,j|E_0).
\end{equation}
Since the band is away from the ground state, the particles will be exponentially damped, irrespective of the fact that the states in the band are, according to this independent particle picture, delocalized.

So at the transition the picture is that of a finite density of particles which interact and annihilate in couples (like anyons) and whose gain in energy due to delocalization makes the vacuum state unstable.

As the single particle eigenstates are all delocalized, the only possible source of localization could come from a large susceptibility of the many-body system to disorder, like hypothesized in \cite{schiulaz2013ideal,pino2015non} (but see also \cite{yao2014quasi}).
This point is worth investigating and we will address it in future work, where we anticipate that different numerical techniques will be needed as QMC numerics is not suited to answer this kind of questions.

We now move to the prediction of this mean field for the entanglement entropy of the ground state. Let us use the basis $\ket{0},\ket{i}, \ket{i,j}, \ket{i,j,k},\dotsc$ . Then writing a generic state in this basis
\begin{equation}
\ket{\Psi}=c_0\ket{0}+\sum_{i=1}^Nc_i\ket{i}+\sum_{i,j=1}^Nc_{i,j}\ket{i,j}+\dotsb\ .
\end{equation}
The reduced density matrix $\rho_A=\Tr_B\ket{\Psi}\bra{\Psi}$ can be broken in different blocks pertaining to different particle numbers
\begin{equation}
\rho_A=\rho_0\oplus\rho_1\oplus \dotsb \oplus \rho_{N_A},
\end{equation}
where $N_A$ is the number of sites in $A$, which is the maximum number of particles allowed in the region. Any R\'enyi entropy, including the entanglement entropy, will be written as a sum over the R\'enyi entropies of the different sectors:
\begin{equation}
\Tr\rho_A^2=\sum_{n=1}^{N_A}\Tr\rho_n^2.
\end{equation}
Now, the sector with $n$ particles contains at most $\binom{N_A}{n}$ states, so it is clear that if we want to have
\begin{equation}
S^{(2)}=-\ln\Tr\rho_A^2\propto N_A
\end{equation}
as our numerics shows, we need to have $n\propto N_A$ so that at least one of the contributions $\Tr\rho_n^2\sim e^{s N_A}$ is exponential in $N_A \propto N$. So, both the extensivity of the R\'enyi entropy and the correction to $\Gamma_c$ point in the direction of a finite density of excitations in the ground state.
 
Summarizing this Section, and refraining from doing numerology, there is a lesson to be learned from a mean-field theory of excitations on the $x$-polarized ground state. First of all, the starting point $\Gamma=\infty$ seems a good one, both qualitatively and quantitatively, in particular at large $K$. It is very instructive, in order to build an intuition of this glass transition, to consider this phenomenology complementary to the one obtained by starting from the classical $z$-spin glass at $\Gamma=0$ and the dressing it with excitations. Then again, it seems that looking at different particle numbers sector $n$ we get a better approximation going up with $n$ so the transition occurs when a highly occupied state comes close to the ground state, which has only ``virtual" particles. 

This points in a direction which is completely different from that advocated in previous studies. The phase transition is dominated by quantum processes and probably has little signature of the complexity of the classical problem. Since it is reasonable to expect that the latter has to manifest itself at some point during the adiabatic evolution, it is conceivable that an inner region of the glassy phase is characterized by another phase transition. A natural candidate would be an MBL region, which is typically affected by exponentially small gaps, which would therefore be encountered by the adiabatic algorithm before the classical point. This interesting research direction calls for further work.

\section{Conclusions}\label{sect:conclusions}

We studied the $T = 0$ line of the $(T,\Gamma)$-phase diagram of a transverse-field Ising spin glass defined on a regular random graph of connectivity $K=2$, where the system is known to enter a glassy phase at small values of $\Gamma$. We computed numerically the disorder-averaged R\'enyi entropy $S_A^{(2)}$ when the region $A$ is taken to be half of the system. We focused on the paramagnetic phase up to the critical point. We found that the R\'enyi entanglement entropy satisfies a volume law for all values of the transverse field $\Gamma$ we considered, with a prefactor that is maximal at a point we interpret as the critical point of a QPT. We see that this point coincides within statistical errors with the critical point of the glassy phase transition of the model, which we found by studying the Edwards--Anderson parameter $q_{EA}$. We also saw that $q_{EA}$ is continuous at this critical point.

We studied the disorder-averaged values of two (spatial) two-point correlation functions of the model --- the maximal correlation and the mean correlation --- in order to get a better understanding of this phase transition. The decay of both of these correlation functions was found to be compatible with a stretched-exponential for all values of $\Gamma$, both critical and off-critical. We extracted critical correlation lengths $\xi_N(\Gamma_c)$ and observed that they decrease with the system size $N$ and converge to a finite value at the critical point of the glassy transition. We conclude that this phase transition exhibits features of both first-order transitions (finite critical correlation length) and second-order ones (continuous order parameter). From all the above-mentioned quantities we extracted the following estimate for the critical point: $1.82 \leq \Gamma_c \leq 1.85$.

Small-size numerical studies of the quantum Fisher information point to the fact that critical and off-critical multipartite entanglement are microscopic, \ie they never involve more than a number of spins that is constant in the system size. This might \emph{prima facie} seem to contrast with the volume-law of the R\'enyi entropy but we believe that the following simple picture of the wavefunction to be compatible with all the data. At $\Gamma = \infty$ the wavefunction is a product state of $\sigma^x$ eigenstates. As $\Gamma$ is decreased, pairs of nearest-neighbours entangled spins are created and become increasingly entangled as $\Gamma$ approaches the critical point of the QPT. The volume-law scaling of the R\'enyi entropy is a consequence of the expander-graph structure of the interaction graph of the model: any bipartition $A$ cuts through an extensive number of these $2$-spin entangled states, each of which gives a finite contribution to the entanglement entropy $S^{(2)}_A$.

In order to further explain these results we attempted a perturbative calculation using the large-$\Gamma$ transverse-field term as the unperturbed Hamiltonian and the spin-glass coupling strength $J$ as the perturbative parameter. The spectrum of the unperturbed Hamiltonian can be interpreted as a vacuum of quasiparticles, on top of which there are equally-spaced bands of increasing quasiparticle density. First-order parturbation theory in $J$ showed that the energy of the first excited state (belonging to the one-particle band) crosses the ground state energy at $\Gamma_c \approx 1.414$. Going to higher bands and using a mean-field Hamiltonian that reproduces some of the contributions of higher-order perturbation terms we see that the critical points moves to higher values of $\Gamma$. The idea is then that at the critical point one of the higher bands crosses the vacuum energy of the quasiparticles, so that energy considerations favour the creation of a finite density of quasiparticles. This perturbative approach is also corroborated by the fact that its predictions are consistent with the known results for the limit $K \rightarrow \infty$ of infinite connectivity, where the RRG Ising spin glass goes to the Sherrington-Kirkpatrick model.

From the point of view of quantum computation we note that the microscopicity of the entanglement, along with the fact that exponentially-closing avoided crossings in the paramagnetic region are highly unlikely, suggest on the one hand that quantum annealers do not need to generate and sustain an inordinate amount of entanglement in order to correctly follow the adiabatic path, at least up to the critical point of the transition. On the other hand, this part of the adiabatic-evolution dynamics can be efficiently simulated on a classical computer using the standard methods of Refs \cite{vanDam2001,jozsa2011}. The extension of these considerations to the entire adiabatic path rests upon the scaling behaviour of the multipartite entanglement and the minimal gap inside of the glassy phase. Previous results on the related problem $3$-regular MAX-CUT (which is equivalent to our model without disordered interactions $J_{ij}$) found that the gap closes superpolynomially fast inside of the glassy phase. We expect this to be the same also in the Ising spin glass. Also, classical considerations suggest that at least for $\Gamma = 0$ we should have ground states that are superpositions of (quantum states that represent) the solutions to the classical problem, \ie the classical Ising spin glass. These quantum states are expected to be products of smaller entangled and unentangled states. For example, if ``000101'' and ``000011'' are the only two solutions to the classical problem, then their superposition can be written as
\[
 \ket{000101} + \ket{000011} = \ket{000} \otimes (\ket{10}+\ket{01}) \otimes \ket{1}
\]
that is, a six-qubit state that is only $2$-entangled. Therefore entanglement is dependent on the structure of the solutions to the classical problem, for which at the moment we do not have a clear picture.

{We believe there are several future directions that might be worth exploring. The obvious one is to extend our study of the R\'enyi entropy inside of the glassy phase, where ergodicity breaking manifests itself as a fast increase of the convergence time needed by the Monte Carlo simulations. Methods such as parallel tempering have been used to ameliorate this effect so it is reasonable to believe that this obstacle is not an insurmountable one. Secondly, a better understanding of the entanglement in the ground-state wavefunctions of the Quantum Adiabatic Algorithm seems auspicable. In particular, when does an adiabatic path encounters points of macroscopic entanglement, and how does this feature relate with the minimal gap? Still another possible direction is to study the R\'enyi entropy dynamics in the adiabatic path of a realistic combinatorial problem. Following the literature, the average entanglement properties of $k$-SAT or EXACT COVER in a transverse field might be interesting points. Finally, a more systematic study of the mean-field approach and of the developement of quasiparticle descriptions of the physics of adiabatic paths seem useful in order to develop a better intuition.}
\section{Acknowledgements}
We would like to thank G. Parisi and F. Ricci-Tersenghi for discussions.

%

\appendix

\section{Quantum Monte Carlo, detailed discussion}

In order to compute the thermodynamic properties of the Ising models with transverse field defined by eq.~(\ref{eq:tfisg}) we adopt the path-integral formalism within the primitive approximation~\cite{suzuki1,suzuki2}. 
This allows us to approximate the  canonical partition function $Z$ of the quantum model with the (approximately) equivalent partition function $Z_c$ of a classical system with an additional dimension, corresponding to the imaginary-time direction. 
Formally, one has:
\begin{equation}
Z=\Tr\left(\rho\right)\simeq Z_c = C^{NM}\sum_{s^1,\dotsc,s^M} \exp\left(-\beta H_c\right) ,
\end{equation}
where $\rho=\exp\left(-\beta H\right)$ is the density matrix, its trace is $\Tr\left(\rho\right)=\sum_{s}\left< s \left| \exp\left(-\beta H\right) \right|s \right>$ (being $\left|s\right>$ our computational basis), and the Hamiltonian of the effective classical model is 
\begin{equation}
H_c = -\sum_{m=1}^{M} \left(  \sum_{\left<i,j\right>} J_{ij} s_i^m s_j^m + J_{\perp} \sum_{i=1}^{N} s_i^m s_i^{m+1} \right) ;
\end{equation}
here, the Trotter (integer) number $M$ corresponds to the number of time slices, $s_i^m=\pm1$ is the spin variable $i$ (with $i=1,\dotsc,N$) at the time slice $m$ (with $m=1,\dotsc,M$), $s^m=\left\{s_1^m,\dotsc,s_N^m\right\}$ denotes the spin configuration at the slice $m$, the periodic boundary condition $s^{M+1}=s^1$ follows from the trace operation in the definition of $Z$, $J_{\perp} =-(MT/2)\ln \tanh \left( \Gamma /MT\right) > 0$ is the strength of a ferromagnetic coupling between corresponding spins at consecutive time slices, and the normalization constant is $C = \left((1/2)\sinh\left(2\Gamma/MT\right)\right)^{1/2}$. 
The above approximation depicts a systematic error proportional to the square of the time-step $\Delta \tau = \beta/M$, which vanishes in the large $M$ limit.
Thanks to this quantum-to-classical mapping, the static properties of the system, including the thermodynamic potentials and the correlation functions, can be computed via standard Monte Carlo techniques by sampling (e.g., using the Metropolis algorithm) configurations according to the probability density function $p(s^1,\dotsc, s^{M}) = C^{NM}\exp\left(-\beta H_c\right)/Z_c$. 
This approach has also been employed in simulations of the quantum annealing of quantum Ising glasses~\cite{car,martovnak}, and in the calculation of minimum gaps along adiabatic annealing dynamics~\cite{young}
However, the computation of entropies, and in particular of the R\'{e}nyi entanglement entropy, is a challenging computational task, even when the Monte Carlo simulations of the effective classical model can be efficiently performed without negative sign problems nor frustration.\\

\begin{figure}[htbp]
\begin{center}
\includegraphics[width=\columnwidth]{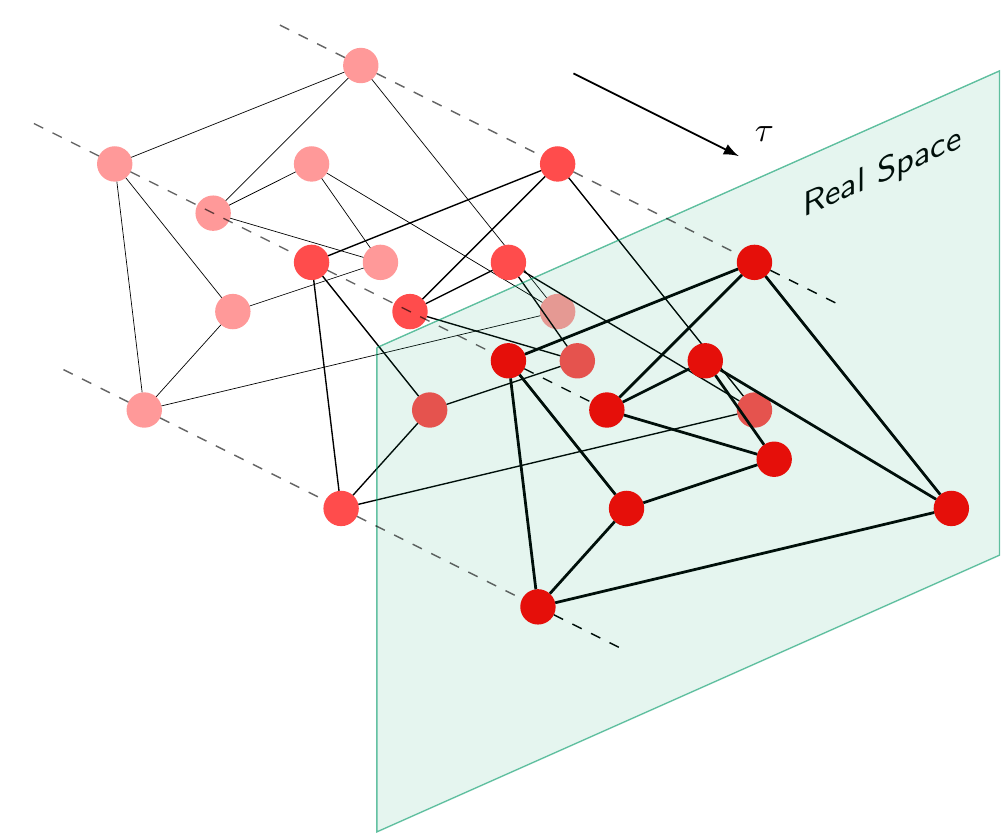}
\caption{A quantum theory on a cubic graph is turned via the quantum-to-classical mapping into an equivalent classical theory on a 5-regular graph, where an additional “imaginary time” dimension is to be taken into account. Effective couplings along this directions, some of which are represented by dashed lines in the above figure, replace the real-space transverse interactions of the quantum theory.}
\label{fig:qmc}
\includegraphics[width=\columnwidth]{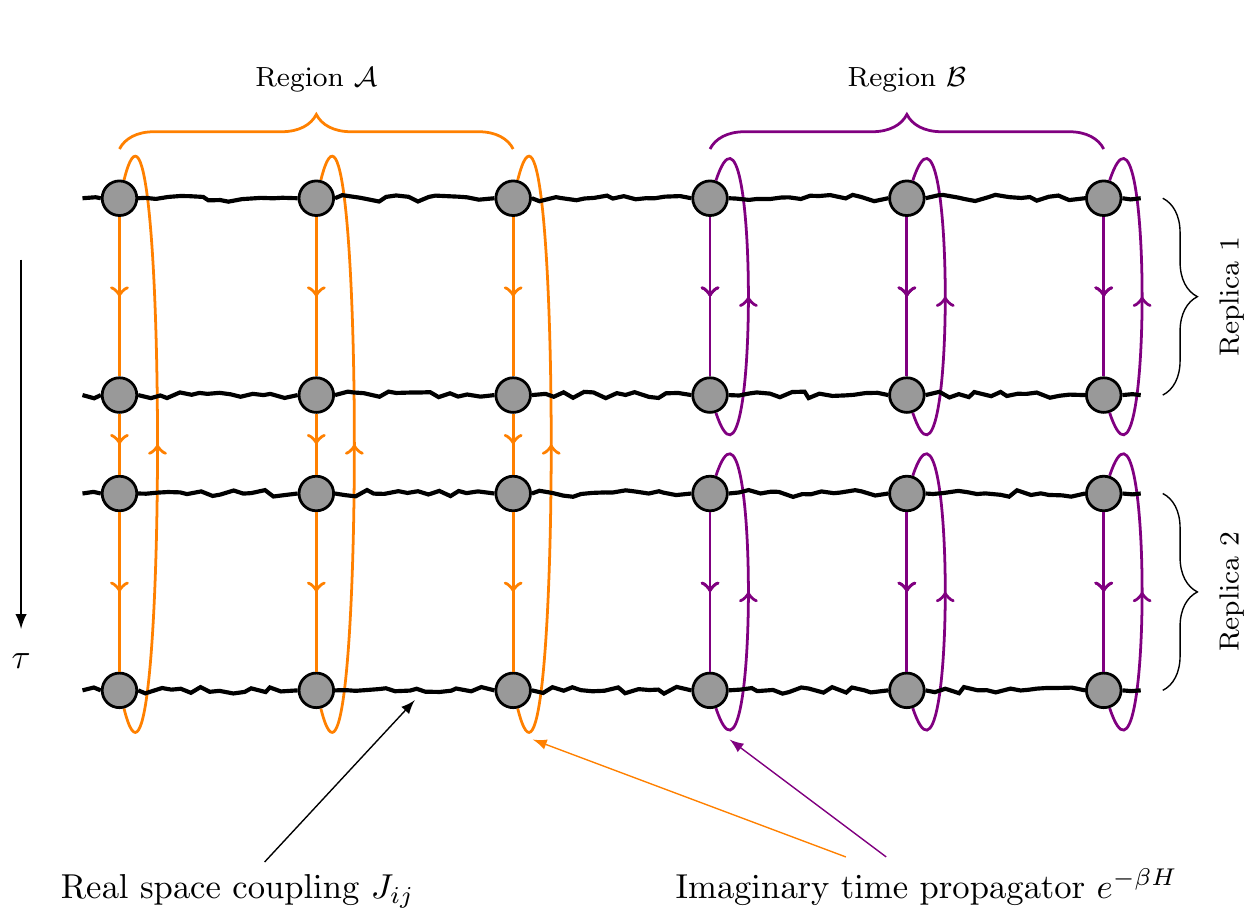}
\caption{Pictorial representation of the system described by $Z^{(2)}_A$. Two copies of the original system are connected by imaginary time couplings, but only in the region $A$. Wobbly lines denote random interactions, solid lines are ferromagnetic.}
\label{fig:replicas}
\end{center}
\end{figure}

Several approaches to compute R\'{e}nyi entropies using Monte Carlo simulations have been engineered, including the temperature-integration method of Ref.~\cite{melko}, the swap operator method of Ref.~\cite{hastings} (for $SU(2)$-invariant lattice spin systems), the weight-ratio estimator method of Ref.~\cite{caraglio,gliozzi,alba}, the mixed-ensemble method of Ref.~\cite{buividovich}, and  the extended configuration-space method of Ref~\cite{roscilde}. 
While all these methods have their own appealing features, we adopt the extended configuration-space method, since it is extremely versatile and, more importantly, efficient in the large system-size regime. We briefly describe it below, following Ref~\cite{roscilde}.\\
 This method is based on the identity derived in Ref.~\cite{calabrese}:
 \begin{equation}
 S_A^{(\alpha)} = \frac{\log\left(R_A^{(\alpha)}  \right)}{1-\alpha},
 \end{equation}
where $R_A^{(\alpha)} = Z_A^{(\alpha)} / Z^\alpha$ is the ratio between the partition function corresponding to $\alpha$ equivalent replicas of the system $Z^\alpha=\left[ \Tr\left(\rho\right)\right]^\alpha$, and the partition function of the $\alpha$ replicas with the corresponding $a$ subsystem cyclically interconnected, which we denote as $Z_A^{(\alpha)}$. What this means is more easily explained by considering the specific instance $\alpha=2$, which is the case treated in this Article. 
In this case, one has:
\begin{equation}
Z_A^{(2)} = \sum_{s_A,s_B,\tilde{s}_A,\tilde{s}_B} \left< s_As_B \left| \rho \right| \tilde{s}_A s_B \right>  \left< \tilde{s}_A\tilde{s}_B \left| \rho \right| s_A \tilde{s}_B \right> ,
\end{equation}
where $\{ \ket{s_As_B} \}$ is the basis set with the spin degrees of freedom $s_A$ and $s_B$ (corresponding, respectively, to the complementary regions $A$ and $B$ of a single replica) separately indicated, and where $s_A,s_B$ and $\tilde{s}_A,\tilde{s}_B$ indicate spins in the first and in the second replica, respectively. One can apply to $Z^2$ and to $Z_A^{(2)}$ the path-integral formalism explained above in the case of $Z$. The sampling probability corresponding to $Z^2$ is simply the product $p^{(2)}(\Sigma)=p(s^1,\dotsc,s^M)p(\tilde{s}^1,\dotsc,\tilde{s}^M)$ (where $\Sigma=\left\{s^1,\dotsc,s^M,\tilde{s}^1,\dotsc,\tilde{s}^M\right\}$ is the spin configuration of the combined system);  the probability $p_A^{(2)}(\Sigma)$, corresponding to $Z_A^{(2)}$, is also obtained in a straightforward manner by employing the modified imaginary-time boundary conditions:
$s_i^{M+1}=\tilde{s}_i^{1}$ and $\tilde{s}_i^{M+1}=s_i^{1}$ if $i\in A$, and $s_i^{M+1}=s_i^{1}$ and $\tilde{s}_i^{M+1}=\tilde{s}_i^{1}$ if $i\in B$.\\
The computation of the ratio $R_A^{(2)}$ can implemented following an approach analogous to the the worm-algorithm simulations~\cite{tupitsyn}, in which one computes properties that are off-diagonal in the computational basis by sampling an extended configuration space.
In the case of interest here, one has to sample the extended configuration space corresponding to the union of partition functions $Z^2 \bigcup Z_A^{(2)}$. This includes configurations corresponding to $Z^2$, which are sampled according to the probability $p^{(2)}(\Sigma)$, and those corresponding to  $Z_A^{(2)}$, sampled according to $p_A^{(2)}(\Sigma)$. Beyond the standard Metropolis updates which drive the Monte Carlo dynamics within the two sectors $Z^2$ and $Z_A^{(2)}$, separately, one includes updates that propose jumps from the $Z^2$-sector to the $Z_A^{(2)}$-sector, and vice versa. The formers are accepted with the probability $P_A=\min\left(1,p_A^{(2)}(\Sigma)/p^{(2)}(\Sigma)\right)$, the latter with probability $P_{A=\emptyset}=\min\left(1,p^{(2)}(\Sigma)/p_a^{(2)}(\Sigma)\right)$. Its worth noticing that these acceptance probabilities only depend on the links at the imaginary times $m=1$ and $m=M$ in the $A$ subsystem of the two replicas. Attention has to be paid in proposing the two opposite updates with balanced frequencies.\\
Within this extended Monte Carlo simulation, the ratio $R_A^{(2)}$ can be estimated computing the Monte Carlo average $R_A^{(2)}=\left< N_A/N_{A=\emptyset}\right>_{\mathrm{MC}}$, where the integer $N_A$ counts how many times the system is found in the $Z_A^{(2)}$ sector during the simulations, and $N_{A=\emptyset}$ counts how many times it is found in the $Z^{2}$ sector.\\
It is worth mentioning that the efficiency of this algorithm might degrade when the subsystem size $l_A$ increases, due to the decrease of the acceptance rate of the Monte Carlo updates that switch between the two sectors $Z^{2}$ and $Z^{(2)}_A$, resulting in large statistical fluctuations. This eventual problem could be circumvented by using the incremental formula implemented in Refs.~\cite{melko,roscilde}, where the R\'{e}nyi entropy of a (large) subsystem $A$ is obtained by decomposing it into subparts of increasing size; however, we verified that in the disordered RRG considered in this work, the basic algorithm (without the incremental decomposition) is sufficiently efficient even for considerably large system sizes, and that the incremental formula does not provide a critical performance boost.

 \section{QFI details}
 
 We have that \cite{hyllus2012}, for a pure state $\ket{\psi}$ and $\hat{O} = \frac{1}{2} \sum_i \vec{n_i} \cdot \vec{\sigma}$, the quantum Fisher information takes a particularly simple form
 
 \[
 	F_Q[\psi,\hat{O}] = 4 \Big(\avg{\psi \rvert \hat{O}\hat{O} \lvert \psi} - \avg{\psi \rvert \hat{O} \lvert \psi}^2\Big)
 \]
 
 Now, it's easy to see that, \eg for $\hat{O} \equiv \frac{1}{2}J_z$

 \begin{eqnarray*}
 F[\psi_0; J_z] &=& \sum_{i=1}^N \avg{\psi_0 \rvert \sigma_i^z\sigma_i^z \lvert \psi_0}-\avg{\psi_0 \rvert \sigma_i^z \lvert \psi_0}^2\\
 &+& \sum_{i\neq j} \avg{\psi_0 \rvert \sigma_i^z \sigma_j^z \lvert \psi_0} - \avg{\psi_0 \rvert \sigma_i^z \lvert \psi_0}\avg{\psi_0 \rvert \sigma_j^z \lvert \psi_0} \\
 			&=& N(1-q_{EA}) + 2\sum_{i < j} C_{ij}
 \end{eqnarray*}
 where $q_{EA}$ and $C_{ij}$ are the Edwards--Anderson order parameter and the correlation funcions defined in (respectively) Eq. (\ref{eq:eaop}) and Eq. (\ref{eq:correlation_function}). Similar formulas hold for $J_y$ and $J_x$.

 The QFI density $f \equiv F/N$ is then given by

 \[
 f[\psi_0,J_z] = (1-q_{EA}) + \frac{2}{N} \sum_{i<j} C_{ij}
 \]
 ergo the glassy behaviour is never an indicator of quantum macroscopicity (as defined in \cite{macroscopicity}, \ie $f_Q > O(1)$) and actually, glassy behaviour seems to \emph{suppress} multipartite entanglement. The only possible contribution is from the $O(N^2)$ two-point correlation terms that need to give a superlinear scaling.

Now, suppose the correlation functions $C_{ij}$ depend only on the distance $r = \lvert i - j \rvert$ between the spins. Then one can rewrite the second term as
\[
	\sum_{r=1}^{r_{\max}} N_r C(r)
\]
where $N_r$ is the number of pairs of spins $(i,j)$ in the system that are at distance $r$.


\begin{thebibliography}{75}%
\makeatletter
\providecommand \@ifxundefined [1]{%
 \@ifx{#1\undefined}
}%
\providecommand \@ifnum [1]{%
 \ifnum #1\expandafter \@firstoftwo
 \else \expandafter \@secondoftwo
 \fi
}%
\providecommand \@ifx [1]{%
 \ifx #1\expandafter \@firstoftwo
 \else \expandafter \@secondoftwo
 \fi
}%
\providecommand \natexlab [1]{#1}%
\providecommand \enquote  [1]{``#1''}%
\providecommand \bibnamefont  [1]{#1}%
\providecommand \bibfnamefont [1]{#1}%
\providecommand \citenamefont [1]{#1}%
\providecommand \href@noop [0]{\@secondoftwo}%
\providecommand \href [0]{\begingroup \@sanitize@url \@href}%
\providecommand \@href[1]{\@@startlink{#1}\@@href}%
\providecommand \@@href[1]{\endgroup#1\@@endlink}%
\providecommand \@sanitize@url [0]{\catcode `\\12\catcode `\$12\catcode
  `\&12\catcode `\#12\catcode `\^12\catcode `\_12\catcode `\%12\relax}%
\providecommand \@@startlink[1]{}%
\providecommand \@@endlink[0]{}%
\providecommand \url  [0]{\begingroup\@sanitize@url \@url }%
\providecommand \@url [1]{\endgroup\@href {#1}{\urlprefix }}%
\providecommand \urlprefix  [0]{URL }%
\providecommand \Eprint [0]{\href }%
\providecommand \doibase [0]{http://dx.doi.org/}%
\providecommand \selectlanguage [0]{\@gobble}%
\providecommand \bibinfo  [0]{\@secondoftwo}%
\providecommand \bibfield  [0]{\@secondoftwo}%
\providecommand \translation [1]{[#1]}%
\providecommand \BibitemOpen [0]{}%
\providecommand \bibitemStop [0]{}%
\providecommand \bibitemNoStop [0]{.\EOS\space}%
\providecommand \EOS [0]{\spacefactor3000\relax}%
\providecommand \BibitemShut  [1]{\csname bibitem#1\endcsname}%
\let\auto@bib@innerbib\@empty
\bibitem [{\citenamefont {Farhi}\ \emph {et~al.}(2001)\citenamefont {Farhi},
  \citenamefont {Goldstone}, \citenamefont {Gutmann}, \citenamefont {Lapan},
  \citenamefont {Lundgren},\ and\ \citenamefont {Preda}}]{farhi2001quantum}%
  \BibitemOpen
  \bibfield  {author} {\bibinfo {author} {\bibfnamefont {E.}~\bibnamefont
  {Farhi}}, \bibinfo {author} {\bibfnamefont {J.}~\bibnamefont {Goldstone}},
  \bibinfo {author} {\bibfnamefont {S.}~\bibnamefont {Gutmann}}, \bibinfo
  {author} {\bibfnamefont {J.}~\bibnamefont {Lapan}}, \bibinfo {author}
  {\bibfnamefont {A.}~\bibnamefont {Lundgren}}, \ and\ \bibinfo {author}
  {\bibfnamefont {D.}~\bibnamefont {Preda}},\ }\href@noop {} {\bibfield
  {journal} {\bibinfo  {journal} {Science}\ }\textbf {\bibinfo {volume}
  {292}},\ \bibinfo {pages} {472} (\bibinfo {year} {2001})}\BibitemShut
  {NoStop}%
\bibitem [{\citenamefont {Kadowaki}\ and\ \citenamefont
  {Nishimori}(1998)}]{kadowaki98quantum}%
  \BibitemOpen
  \bibfield  {author} {\bibinfo {author} {\bibfnamefont {T.}~\bibnamefont
  {Kadowaki}}\ and\ \bibinfo {author} {\bibfnamefont {H.}~\bibnamefont
  {Nishimori}},\ }\href {\doibase 10.1103/PhysRevE.58.5355} {\bibfield
  {journal} {\bibinfo  {journal} {Phys. Rev. E}\ }\textbf {\bibinfo {volume}
  {58}},\ \bibinfo {pages} {5355} (\bibinfo {year} {1998})}\BibitemShut
  {NoStop}%
\bibitem [{\citenamefont {Boixo}\ \emph {et~al.}(2014)\citenamefont {Boixo},
  \citenamefont {R{\o}nnow}, \citenamefont {Isakov}, \citenamefont {Wang},
  \citenamefont {Wecker}, \citenamefont {Lidar}, \citenamefont {Martinis},\
  and\ \citenamefont {Troyer}}]{boixo2014evidence}%
  \BibitemOpen
  \bibfield  {author} {\bibinfo {author} {\bibfnamefont {S.}~\bibnamefont
  {Boixo}}, \bibinfo {author} {\bibfnamefont {T.~F.}\ \bibnamefont
  {R{\o}nnow}}, \bibinfo {author} {\bibfnamefont {S.~V.}\ \bibnamefont
  {Isakov}}, \bibinfo {author} {\bibfnamefont {Z.}~\bibnamefont {Wang}},
  \bibinfo {author} {\bibfnamefont {D.}~\bibnamefont {Wecker}}, \bibinfo
  {author} {\bibfnamefont {D.~A.}\ \bibnamefont {Lidar}}, \bibinfo {author}
  {\bibfnamefont {J.~M.}\ \bibnamefont {Martinis}}, \ and\ \bibinfo {author}
  {\bibfnamefont {M.}~\bibnamefont {Troyer}},\ }\href@noop {} {\bibfield
  {journal} {\bibinfo  {journal} {Nature Physics}\ }\textbf {\bibinfo {volume}
  {10}},\ \bibinfo {pages} {218} (\bibinfo {year} {2014})}\BibitemShut
  {NoStop}%
\bibitem [{\citenamefont {Laumann}\ \emph {et~al.}(2015)\citenamefont
  {Laumann}, \citenamefont {Moessner}, \citenamefont {Scardicchio},\ and\
  \citenamefont {Sondhi}}]{laumann2015quantum}%
  \BibitemOpen
  \bibfield  {author} {\bibinfo {author} {\bibfnamefont {C.~R.}\ \bibnamefont
  {Laumann}}, \bibinfo {author} {\bibfnamefont {R.}~\bibnamefont {Moessner}},
  \bibinfo {author} {\bibfnamefont {A.}~\bibnamefont {Scardicchio}}, \ and\
  \bibinfo {author} {\bibfnamefont {S.}~\bibnamefont {Sondhi}},\ }\href@noop {}
  {\bibfield  {journal} {\bibinfo  {journal} {The European Physical Journal
  Special Topics}\ }\textbf {\bibinfo {volume} {224}},\ \bibinfo {pages} {75}
  (\bibinfo {year} {2015})}\BibitemShut {NoStop}%
\bibitem [{\citenamefont {Kirkpatrick}\ \emph {et~al.}(1983)\citenamefont
  {Kirkpatrick}, \citenamefont {Vecchi} \emph
  {et~al.}}]{kirkpatrick1983optimization}%
  \BibitemOpen
  \bibfield  {author} {\bibinfo {author} {\bibfnamefont {S.}~\bibnamefont
  {Kirkpatrick}}, \bibinfo {author} {\bibfnamefont {M.~P.}\ \bibnamefont
  {Vecchi}},  \emph {et~al.},\ }\href@noop {} {\bibfield  {journal} {\bibinfo
  {journal} {science}\ }\textbf {\bibinfo {volume} {220}},\ \bibinfo {pages}
  {671} (\bibinfo {year} {1983})}\BibitemShut {NoStop}%
\bibitem [{\citenamefont {Kirkpatrick}\ and\ \citenamefont
  {Selman}(1994)}]{kirkpatrick1994critical}%
  \BibitemOpen
  \bibfield  {author} {\bibinfo {author} {\bibfnamefont {S.}~\bibnamefont
  {Kirkpatrick}}\ and\ \bibinfo {author} {\bibfnamefont {B.}~\bibnamefont
  {Selman}},\ }\href@noop {} {\bibfield  {journal} {\bibinfo  {journal}
  {Science}\ }\textbf {\bibinfo {volume} {264}},\ \bibinfo {pages} {1297}
  (\bibinfo {year} {1994})}\BibitemShut {NoStop}%
\bibitem [{\citenamefont {Monasson}\ \emph {et~al.}(1999)\citenamefont
  {Monasson}, \citenamefont {Zecchina}, \citenamefont {Kirkpatrick},
  \citenamefont {Selman},\ and\ \citenamefont
  {Troyansky}}]{monasson1999determining}%
  \BibitemOpen
  \bibfield  {author} {\bibinfo {author} {\bibfnamefont {R.}~\bibnamefont
  {Monasson}}, \bibinfo {author} {\bibfnamefont {R.}~\bibnamefont {Zecchina}},
  \bibinfo {author} {\bibfnamefont {S.}~\bibnamefont {Kirkpatrick}}, \bibinfo
  {author} {\bibfnamefont {B.}~\bibnamefont {Selman}}, \ and\ \bibinfo {author}
  {\bibfnamefont {L.}~\bibnamefont {Troyansky}},\ }\href@noop {} {\bibfield
  {journal} {\bibinfo  {journal} {Nature}\ }\textbf {\bibinfo {volume} {400}},\
  \bibinfo {pages} {133} (\bibinfo {year} {1999})}\BibitemShut {NoStop}%
\bibitem [{\citenamefont {M{\'e}zard}\ \emph {et~al.}(2002)\citenamefont
  {M{\'e}zard}, \citenamefont {Parisi},\ and\ \citenamefont
  {Zecchina}}]{mezard2002analytic}%
  \BibitemOpen
  \bibfield  {author} {\bibinfo {author} {\bibfnamefont {M.}~\bibnamefont
  {M{\'e}zard}}, \bibinfo {author} {\bibfnamefont {G.}~\bibnamefont {Parisi}},
  \ and\ \bibinfo {author} {\bibfnamefont {R.}~\bibnamefont {Zecchina}},\
  }\href@noop {} {\bibfield  {journal} {\bibinfo  {journal} {Science}\ }\textbf
  {\bibinfo {volume} {297}},\ \bibinfo {pages} {812} (\bibinfo {year}
  {2002})}\BibitemShut {NoStop}%
\bibitem [{\citenamefont {Bray}\ and\ \citenamefont
  {Moore}(1980)}]{bray1980replica}%
  \BibitemOpen
  \bibfield  {author} {\bibinfo {author} {\bibfnamefont {A.}~\bibnamefont
  {Bray}}\ and\ \bibinfo {author} {\bibfnamefont {M.}~\bibnamefont {Moore}},\
  }\href@noop {} {\bibfield  {journal} {\bibinfo  {journal} {Journal of Physics
  C: Solid State Physics}\ }\textbf {\bibinfo {volume} {13}},\ \bibinfo {pages}
  {L655} (\bibinfo {year} {1980})}\BibitemShut {NoStop}%
\bibitem [{\citenamefont {Goldschmidt}\ and\ \citenamefont
  {Lai}(1990)}]{goldschmidt1990ising}%
  \BibitemOpen
  \bibfield  {author} {\bibinfo {author} {\bibfnamefont {Y.~Y.}\ \bibnamefont
  {Goldschmidt}}\ and\ \bibinfo {author} {\bibfnamefont {P.-Y.}\ \bibnamefont
  {Lai}},\ }\href@noop {} {\bibfield  {journal} {\bibinfo  {journal} {Physical
  review letters}\ }\textbf {\bibinfo {volume} {64}},\ \bibinfo {pages} {2467}
  (\bibinfo {year} {1990})}\BibitemShut {NoStop}%
\bibitem [{\citenamefont {Buccheri}\ \emph {et~al.}(2011)\citenamefont
  {Buccheri}, \citenamefont {De~Luca},\ and\ \citenamefont
  {Scardicchio}}]{buccheri2011structure}%
  \BibitemOpen
  \bibfield  {author} {\bibinfo {author} {\bibfnamefont {F.}~\bibnamefont
  {Buccheri}}, \bibinfo {author} {\bibfnamefont {A.}~\bibnamefont {De~Luca}}, \
  and\ \bibinfo {author} {\bibfnamefont {A.}~\bibnamefont {Scardicchio}},\
  }\href@noop {} {\bibfield  {journal} {\bibinfo  {journal} {Physical Review
  B}\ }\textbf {\bibinfo {volume} {84}},\ \bibinfo {pages} {094203} (\bibinfo
  {year} {2011})}\BibitemShut {NoStop}%
\bibitem [{\citenamefont {Andreanov}\ and\ \citenamefont
  {M{\"u}ller}(2012)}]{andreanov2012long}%
  \BibitemOpen
  \bibfield  {author} {\bibinfo {author} {\bibfnamefont {A.}~\bibnamefont
  {Andreanov}}\ and\ \bibinfo {author} {\bibfnamefont {M.}~\bibnamefont
  {M{\"u}ller}},\ }\href@noop {} {\bibfield  {journal} {\bibinfo  {journal}
  {Physical review letters}\ }\textbf {\bibinfo {volume} {109}},\ \bibinfo
  {pages} {177201} (\bibinfo {year} {2012})}\BibitemShut {NoStop}%
\bibitem [{\citenamefont {Laumann}\ \emph
  {et~al.}(2010{\natexlab{a}})\citenamefont {Laumann}, \citenamefont
  {Moessner}, \citenamefont {Scardicchio},\ and\ \citenamefont
  {Sondhi}}]{laumann2010random}%
  \BibitemOpen
  \bibfield  {author} {\bibinfo {author} {\bibfnamefont {C.~R.}\ \bibnamefont
  {Laumann}}, \bibinfo {author} {\bibfnamefont {R.}~\bibnamefont {Moessner}},
  \bibinfo {author} {\bibfnamefont {A.}~\bibnamefont {Scardicchio}}, \ and\
  \bibinfo {author} {\bibfnamefont {S.}~\bibnamefont {Sondhi}},\ }\href@noop {}
  {\bibfield  {journal} {\bibinfo  {journal} {Quantum Information \&
  Computation}\ }\textbf {\bibinfo {volume} {10}},\ \bibinfo {pages} {1}
  (\bibinfo {year} {2010}{\natexlab{a}})}\BibitemShut {NoStop}%
\bibitem [{\citenamefont {Laumann}\ \emph
  {et~al.}(2010{\natexlab{b}})\citenamefont {Laumann}, \citenamefont
  {L{\"a}uchli}, \citenamefont {Moessner}, \citenamefont {Scardicchio},\ and\
  \citenamefont {Sondhi}}]{laumann2010product}%
  \BibitemOpen
  \bibfield  {author} {\bibinfo {author} {\bibfnamefont {C.~R.}\ \bibnamefont
  {Laumann}}, \bibinfo {author} {\bibfnamefont {A.}~\bibnamefont
  {L{\"a}uchli}}, \bibinfo {author} {\bibfnamefont {R.}~\bibnamefont
  {Moessner}}, \bibinfo {author} {\bibfnamefont {A.}~\bibnamefont
  {Scardicchio}}, \ and\ \bibinfo {author} {\bibfnamefont {S.}~\bibnamefont
  {Sondhi}},\ }\href@noop {} {\bibfield  {journal} {\bibinfo  {journal}
  {Physical Review A}\ }\textbf {\bibinfo {volume} {81}},\ \bibinfo {pages}
  {062345} (\bibinfo {year} {2010}{\natexlab{b}})}\BibitemShut {NoStop}%
\bibitem [{\citenamefont {Sherrington}\ and\ \citenamefont
  {Kirkpatrick}(1975)}]{sherrington1975solvable}%
  \BibitemOpen
  \bibfield  {author} {\bibinfo {author} {\bibfnamefont {D.}~\bibnamefont
  {Sherrington}}\ and\ \bibinfo {author} {\bibfnamefont {S.}~\bibnamefont
  {Kirkpatrick}},\ }\href@noop {} {\bibfield  {journal} {\bibinfo  {journal}
  {Physical review letters}\ }\textbf {\bibinfo {volume} {35}},\ \bibinfo
  {pages} {1792} (\bibinfo {year} {1975})}\BibitemShut {NoStop}%
\bibitem [{\citenamefont {Thouless}\ \emph {et~al.}(1977)\citenamefont
  {Thouless}, \citenamefont {Anderson},\ and\ \citenamefont
  {Palmer}}]{thouless1977solution}%
  \BibitemOpen
  \bibfield  {author} {\bibinfo {author} {\bibfnamefont {D.~J.}\ \bibnamefont
  {Thouless}}, \bibinfo {author} {\bibfnamefont {P.~W.}\ \bibnamefont
  {Anderson}}, \ and\ \bibinfo {author} {\bibfnamefont {R.~G.}\ \bibnamefont
  {Palmer}},\ }\href@noop {} {\bibfield  {journal} {\bibinfo  {journal}
  {Philosophical Magazine}\ }\textbf {\bibinfo {volume} {35}},\ \bibinfo
  {pages} {593} (\bibinfo {year} {1977})}\BibitemShut {NoStop}%
\bibitem [{\citenamefont {Parisi}(1980)}]{parisi1980order}%
  \BibitemOpen
  \bibfield  {author} {\bibinfo {author} {\bibfnamefont {G.}~\bibnamefont
  {Parisi}},\ }\href@noop {} {\bibfield  {journal} {\bibinfo  {journal}
  {Journal of Physics A: Mathematical and General}\ }\textbf {\bibinfo {volume}
  {13}},\ \bibinfo {pages} {1101} (\bibinfo {year} {1980})}\BibitemShut
  {NoStop}%
\bibitem [{\citenamefont {Thouless}(1986)}]{thouless1986spin}%
  \BibitemOpen
  \bibfield  {author} {\bibinfo {author} {\bibfnamefont {D.}~\bibnamefont
  {Thouless}},\ }\href@noop {} {\bibfield  {journal} {\bibinfo  {journal}
  {Physical review letters}\ }\textbf {\bibinfo {volume} {56}},\ \bibinfo
  {pages} {1082} (\bibinfo {year} {1986})}\BibitemShut {NoStop}%
\bibitem [{\citenamefont {M{\'e}zard}\ and\ \citenamefont
  {Parisi}(2001)}]{mezard2001bethe}%
  \BibitemOpen
  \bibfield  {author} {\bibinfo {author} {\bibfnamefont {M.}~\bibnamefont
  {M{\'e}zard}}\ and\ \bibinfo {author} {\bibfnamefont {G.}~\bibnamefont
  {Parisi}},\ }\href@noop {} {\bibfield  {journal} {\bibinfo  {journal} {The
  European Physical Journal B-Condensed Matter and Complex Systems}\ }\textbf
  {\bibinfo {volume} {20}},\ \bibinfo {pages} {217} (\bibinfo {year}
  {2001})}\BibitemShut {NoStop}%
\bibitem [{\citenamefont {Laumann}\ \emph {et~al.}(2008)\citenamefont
  {Laumann}, \citenamefont {Scardicchio},\ and\ \citenamefont
  {Sondhi}}]{laumann2008cavity}%
  \BibitemOpen
  \bibfield  {author} {\bibinfo {author} {\bibfnamefont {C.}~\bibnamefont
  {Laumann}}, \bibinfo {author} {\bibfnamefont {A.}~\bibnamefont
  {Scardicchio}}, \ and\ \bibinfo {author} {\bibfnamefont {S.}~\bibnamefont
  {Sondhi}},\ }\href@noop {} {\bibfield  {journal} {\bibinfo  {journal}
  {Physical Review B}\ }\textbf {\bibinfo {volume} {78}},\ \bibinfo {pages}
  {134424} (\bibinfo {year} {2008})}\BibitemShut {NoStop}%
\bibitem [{\citenamefont {Krzakala}\ \emph {et~al.}(2008)\citenamefont
  {Krzakala}, \citenamefont {Rosso}, \citenamefont {Semerjian},\ and\
  \citenamefont {Zamponi}}]{krzakala2008path}%
  \BibitemOpen
  \bibfield  {author} {\bibinfo {author} {\bibfnamefont {F.}~\bibnamefont
  {Krzakala}}, \bibinfo {author} {\bibfnamefont {A.}~\bibnamefont {Rosso}},
  \bibinfo {author} {\bibfnamefont {G.}~\bibnamefont {Semerjian}}, \ and\
  \bibinfo {author} {\bibfnamefont {F.}~\bibnamefont {Zamponi}},\ }\href@noop
  {} {\bibfield  {journal} {\bibinfo  {journal} {Physical Review B}\ }\textbf
  {\bibinfo {volume} {78}},\ \bibinfo {pages} {134428} (\bibinfo {year}
  {2008})}\BibitemShut {NoStop}%
\bibitem [{\citenamefont {Farhi}\ \emph {et~al.}(2012)\citenamefont {Farhi},
  \citenamefont {Gosset}, \citenamefont {Hen}, \citenamefont {Sandvik},
  \citenamefont {Shor}, \citenamefont {Young},\ and\ \citenamefont
  {Zamponi}}]{farhi2012performance}%
  \BibitemOpen
  \bibfield  {author} {\bibinfo {author} {\bibfnamefont {E.}~\bibnamefont
  {Farhi}}, \bibinfo {author} {\bibfnamefont {D.}~\bibnamefont {Gosset}},
  \bibinfo {author} {\bibfnamefont {I.}~\bibnamefont {Hen}}, \bibinfo {author}
  {\bibfnamefont {A.}~\bibnamefont {Sandvik}}, \bibinfo {author} {\bibfnamefont
  {P.}~\bibnamefont {Shor}}, \bibinfo {author} {\bibfnamefont {A.}~\bibnamefont
  {Young}}, \ and\ \bibinfo {author} {\bibfnamefont {F.}~\bibnamefont
  {Zamponi}},\ }\href@noop {} {\bibfield  {journal} {\bibinfo  {journal}
  {Physical Review A}\ }\textbf {\bibinfo {volume} {86}},\ \bibinfo {pages}
  {052334} (\bibinfo {year} {2012})}\BibitemShut {NoStop}%
\bibitem [{\citenamefont {Basko}\ \emph {et~al.}(2006)\citenamefont {Basko},
  \citenamefont {Aleiner},\ and\ \citenamefont {Altshuler}}]{basko2006metal}%
  \BibitemOpen
  \bibfield  {author} {\bibinfo {author} {\bibfnamefont {D.}~\bibnamefont
  {Basko}}, \bibinfo {author} {\bibfnamefont {I.}~\bibnamefont {Aleiner}}, \
  and\ \bibinfo {author} {\bibfnamefont {B.}~\bibnamefont {Altshuler}},\
  }\href@noop {} {\bibfield  {journal} {\bibinfo  {journal} {Annals of
  physics}\ }\textbf {\bibinfo {volume} {321}},\ \bibinfo {pages} {1126}
  (\bibinfo {year} {2006})}\BibitemShut {NoStop}%
\bibitem [{\citenamefont {Pal}\ and\ \citenamefont {Huse}(2010)}]{pal2010mb}%
  \BibitemOpen
  \bibfield  {author} {\bibinfo {author} {\bibfnamefont {A.}~\bibnamefont
  {Pal}}\ and\ \bibinfo {author} {\bibfnamefont {D.~A.}\ \bibnamefont {Huse}},\
  }\href {\doibase 10.1103/PhysRevB.82.174411} {\bibfield  {journal} {\bibinfo
  {journal} {Phys. Rev. B}\ }\textbf {\bibinfo {volume} {82}},\ \bibinfo
  {pages} {174411} (\bibinfo {year} {2010})}\BibitemShut {NoStop}%
\bibitem [{\citenamefont {Nandkishore}\ and\ \citenamefont
  {Huse}(2014)}]{nandkishore2014many}%
  \BibitemOpen
  \bibfield  {author} {\bibinfo {author} {\bibfnamefont {R.}~\bibnamefont
  {Nandkishore}}\ and\ \bibinfo {author} {\bibfnamefont {D.~A.}\ \bibnamefont
  {Huse}},\ }\href@noop {} {\bibfield  {journal} {\bibinfo  {journal} {arXiv
  preprint arXiv:1404.0686}\ } (\bibinfo {year} {2014})}\BibitemShut {NoStop}%
\bibitem [{\citenamefont {De~Luca}\ and\ \citenamefont
  {Scardicchio}(2013)}]{de2013ergodicity}%
  \BibitemOpen
  \bibfield  {author} {\bibinfo {author} {\bibfnamefont {A.}~\bibnamefont
  {De~Luca}}\ and\ \bibinfo {author} {\bibfnamefont {A.}~\bibnamefont
  {Scardicchio}},\ }\href@noop {} {\bibfield  {journal} {\bibinfo  {journal}
  {Europhys. Lett.}\ }\textbf {\bibinfo {volume} {101}},\ \bibinfo {pages}
  {37003} (\bibinfo {year} {2013})}\BibitemShut {NoStop}%
\bibitem [{\citenamefont {Laumann}\ \emph {et~al.}(2014)\citenamefont
  {Laumann}, \citenamefont {Pal},\ and\ \citenamefont
  {Scardicchio}}]{laumann2014many}%
  \BibitemOpen
  \bibfield  {author} {\bibinfo {author} {\bibfnamefont {C.}~\bibnamefont
  {Laumann}}, \bibinfo {author} {\bibfnamefont {A.}~\bibnamefont {Pal}}, \ and\
  \bibinfo {author} {\bibfnamefont {A.}~\bibnamefont {Scardicchio}},\
  }\href@noop {} {\bibfield  {journal} {\bibinfo  {journal} {Physical review
  letters}\ }\textbf {\bibinfo {volume} {113}},\ \bibinfo {pages} {200405}
  (\bibinfo {year} {2014})}\BibitemShut {NoStop}%
\bibitem [{\citenamefont {Luitz}\ \emph {et~al.}(2015)\citenamefont {Luitz},
  \citenamefont {Laflorencie},\ and\ \citenamefont {Alet}}]{luitz2015many}%
  \BibitemOpen
  \bibfield  {author} {\bibinfo {author} {\bibfnamefont {D.~J.}\ \bibnamefont
  {Luitz}}, \bibinfo {author} {\bibfnamefont {N.}~\bibnamefont {Laflorencie}},
  \ and\ \bibinfo {author} {\bibfnamefont {F.}~\bibnamefont {Alet}},\
  }\href@noop {} {\bibfield  {journal} {\bibinfo  {journal} {Physical Review
  B}\ }\textbf {\bibinfo {volume} {91}},\ \bibinfo {pages} {081103} (\bibinfo
  {year} {2015})}\BibitemShut {NoStop}%
\bibitem [{\citenamefont {Goold}\ \emph {et~al.}(2015)\citenamefont {Goold},
  \citenamefont {Gogolin}, \citenamefont {Clark}, \citenamefont {Eisert},
  \citenamefont {Scardicchio},\ and\ \citenamefont {Silva}}]{goold2015total}%
  \BibitemOpen
  \bibfield  {author} {\bibinfo {author} {\bibfnamefont {J.}~\bibnamefont
  {Goold}}, \bibinfo {author} {\bibfnamefont {C.}~\bibnamefont {Gogolin}},
  \bibinfo {author} {\bibfnamefont {S.}~\bibnamefont {Clark}}, \bibinfo
  {author} {\bibfnamefont {J.}~\bibnamefont {Eisert}}, \bibinfo {author}
  {\bibfnamefont {A.}~\bibnamefont {Scardicchio}}, \ and\ \bibinfo {author}
  {\bibfnamefont {A.}~\bibnamefont {Silva}},\ }\href@noop {} {\bibfield
  {journal} {\bibinfo  {journal} {Physical Review B}\ }\textbf {\bibinfo
  {volume} {92}},\ \bibinfo {pages} {180202} (\bibinfo {year}
  {2015})}\BibitemShut {NoStop}%
\bibitem [{\citenamefont {Lerose}\ \emph {et~al.}(2015)\citenamefont {Lerose},
  \citenamefont {Varma}, \citenamefont {Pietracaprina}, \citenamefont {Goold},\
  and\ \citenamefont {Scardicchio}}]{lerose2015coexistence}%
  \BibitemOpen
  \bibfield  {author} {\bibinfo {author} {\bibfnamefont {A.}~\bibnamefont
  {Lerose}}, \bibinfo {author} {\bibfnamefont {V.~K.}\ \bibnamefont {Varma}},
  \bibinfo {author} {\bibfnamefont {F.}~\bibnamefont {Pietracaprina}}, \bibinfo
  {author} {\bibfnamefont {J.}~\bibnamefont {Goold}}, \ and\ \bibinfo {author}
  {\bibfnamefont {A.}~\bibnamefont {Scardicchio}},\ }\href@noop {} {\bibfield
  {journal} {\bibinfo  {journal} {arXiv preprint arXiv:1511.09144}\ } (\bibinfo
  {year} {2015})}\BibitemShut {NoStop}%
\bibitem [{\citenamefont {Huse}\ \emph {et~al.}(2014)\citenamefont {Huse},
  \citenamefont {Nandkishore},\ and\ \citenamefont
  {Oganesyan}}]{huse2014phenomenology}%
  \BibitemOpen
  \bibfield  {author} {\bibinfo {author} {\bibfnamefont {D.~A.}\ \bibnamefont
  {Huse}}, \bibinfo {author} {\bibfnamefont {R.}~\bibnamefont {Nandkishore}}, \
  and\ \bibinfo {author} {\bibfnamefont {V.}~\bibnamefont {Oganesyan}},\
  }\href@noop {} {\bibfield  {journal} {\bibinfo  {journal} {Physical Review
  B}\ }\textbf {\bibinfo {volume} {90}},\ \bibinfo {pages} {174202} (\bibinfo
  {year} {2014})}\BibitemShut {NoStop}%
\bibitem [{\citenamefont {Imbrie}(2014)}]{imbrie2014many}%
  \BibitemOpen
  \bibfield  {author} {\bibinfo {author} {\bibfnamefont {J.~Z.}\ \bibnamefont
  {Imbrie}},\ }\href@noop {} {\bibfield  {journal} {\bibinfo  {journal} {arXiv
  preprint arXiv:1403.7837}\ } (\bibinfo {year} {2014})}\BibitemShut {NoStop}%
\bibitem [{\citenamefont {Ros}\ \emph {et~al.}(2015)\citenamefont {Ros},
  \citenamefont {Mueller},\ and\ \citenamefont
  {Scardicchio}}]{ros2015integrals}%
  \BibitemOpen
  \bibfield  {author} {\bibinfo {author} {\bibfnamefont {V.}~\bibnamefont
  {Ros}}, \bibinfo {author} {\bibfnamefont {M.}~\bibnamefont {Mueller}}, \ and\
  \bibinfo {author} {\bibfnamefont {A.}~\bibnamefont {Scardicchio}},\
  }\href@noop {} {\bibfield  {journal} {\bibinfo  {journal} {Nuclear Physics
  B}\ }\textbf {\bibinfo {volume} {891}},\ \bibinfo {pages} {420} (\bibinfo
  {year} {2015})}\BibitemShut {NoStop}%
\bibitem [{\citenamefont {Chandran}\ \emph {et~al.}(2015)\citenamefont
  {Chandran}, \citenamefont {Kim}, \citenamefont {Vidal},\ and\ \citenamefont
  {Abanin}}]{chandran2015constructing}%
  \BibitemOpen
  \bibfield  {author} {\bibinfo {author} {\bibfnamefont {A.}~\bibnamefont
  {Chandran}}, \bibinfo {author} {\bibfnamefont {I.~H.}\ \bibnamefont {Kim}},
  \bibinfo {author} {\bibfnamefont {G.}~\bibnamefont {Vidal}}, \ and\ \bibinfo
  {author} {\bibfnamefont {D.~A.}\ \bibnamefont {Abanin}},\ }\href@noop {}
  {\bibfield  {journal} {\bibinfo  {journal} {Physical Review B}\ }\textbf
  {\bibinfo {volume} {91}},\ \bibinfo {pages} {085425} (\bibinfo {year}
  {2015})}\BibitemShut {NoStop}%
\bibitem [{\citenamefont {Altshuler}\ \emph {et~al.}(2010)\citenamefont
  {Altshuler}, \citenamefont {Krovi},\ and\ \citenamefont
  {Roland}}]{altshuler2010anderson}%
  \BibitemOpen
  \bibfield  {author} {\bibinfo {author} {\bibfnamefont {B.}~\bibnamefont
  {Altshuler}}, \bibinfo {author} {\bibfnamefont {H.}~\bibnamefont {Krovi}}, \
  and\ \bibinfo {author} {\bibfnamefont {J.}~\bibnamefont {Roland}},\
  }\href@noop {} {\bibfield  {journal} {\bibinfo  {journal} {Proceedings of the
  National Academy of Sciences}\ }\textbf {\bibinfo {volume} {107}},\ \bibinfo
  {pages} {12446} (\bibinfo {year} {2010})}\BibitemShut {NoStop}%
\bibitem [{\citenamefont {Knysh}\ and\ \citenamefont
  {Smelyanskiy}(2010)}]{knysh2010relevance}%
  \BibitemOpen
  \bibfield  {author} {\bibinfo {author} {\bibfnamefont {S.}~\bibnamefont
  {Knysh}}\ and\ \bibinfo {author} {\bibfnamefont {V.}~\bibnamefont
  {Smelyanskiy}},\ }\href@noop {} {\bibfield  {journal} {\bibinfo  {journal}
  {arXiv preprint arXiv:1005.3011}\ } (\bibinfo {year} {2010})}\BibitemShut
  {NoStop}%
\bibitem [{\citenamefont {Baldwin}\ \emph {et~al.}(2016)\citenamefont
  {Baldwin}, \citenamefont {Laumann}, \citenamefont {Pal},\ and\ \citenamefont
  {Scardicchio}}]{baldwin2016many}%
  \BibitemOpen
  \bibfield  {author} {\bibinfo {author} {\bibfnamefont {C.}~\bibnamefont
  {Baldwin}}, \bibinfo {author} {\bibfnamefont {C.}~\bibnamefont {Laumann}},
  \bibinfo {author} {\bibfnamefont {A.}~\bibnamefont {Pal}}, \ and\ \bibinfo
  {author} {\bibfnamefont {A.}~\bibnamefont {Scardicchio}},\ }\href@noop {}
  {\bibfield  {journal} {\bibinfo  {journal} {Physical Review B}\ }\textbf
  {\bibinfo {volume} {93}},\ \bibinfo {pages} {024202} (\bibinfo {year}
  {2016})}\BibitemShut {NoStop}%
\bibitem [{\citenamefont {{\v{Z}}nidari{\v{c}}}\ \emph
  {et~al.}(2008)\citenamefont {{\v{Z}}nidari{\v{c}}}, \citenamefont {Prosen},\
  and\ \citenamefont {Prelov{\v{s}}ek}}]{vznidarivc2008many}%
  \BibitemOpen
  \bibfield  {author} {\bibinfo {author} {\bibfnamefont {M.}~\bibnamefont
  {{\v{Z}}nidari{\v{c}}}}, \bibinfo {author} {\bibfnamefont {T.}~\bibnamefont
  {Prosen}}, \ and\ \bibinfo {author} {\bibfnamefont {P.}~\bibnamefont
  {Prelov{\v{s}}ek}},\ }\href@noop {} {\bibfield  {journal} {\bibinfo
  {journal} {Physical Review B}\ }\textbf {\bibinfo {volume} {77}},\ \bibinfo
  {pages} {064426} (\bibinfo {year} {2008})}\BibitemShut {NoStop}%
\bibitem [{\citenamefont {Bardarson}\ \emph {et~al.}(2012)\citenamefont
  {Bardarson}, \citenamefont {Pollmann},\ and\ \citenamefont
  {Moore}}]{bardarson2012unbounded}%
  \BibitemOpen
  \bibfield  {author} {\bibinfo {author} {\bibfnamefont {J.~H.}\ \bibnamefont
  {Bardarson}}, \bibinfo {author} {\bibfnamefont {F.}~\bibnamefont {Pollmann}},
  \ and\ \bibinfo {author} {\bibfnamefont {J.~E.}\ \bibnamefont {Moore}},\
  }\href@noop {} {\bibfield  {journal} {\bibinfo  {journal} {Physical review
  letters}\ }\textbf {\bibinfo {volume} {109}},\ \bibinfo {pages} {017202}
  (\bibinfo {year} {2012})}\BibitemShut {NoStop}%
\bibitem [{\citenamefont {De~Almeida}\ and\ \citenamefont
  {Thouless}(1978)}]{de1978stability}%
  \BibitemOpen
  \bibfield  {author} {\bibinfo {author} {\bibfnamefont {J.}~\bibnamefont
  {De~Almeida}}\ and\ \bibinfo {author} {\bibfnamefont {D.~J.}\ \bibnamefont
  {Thouless}},\ }\href@noop {} {\bibfield  {journal} {\bibinfo  {journal}
  {Journal of Physics A: Mathematical and General}\ }\textbf {\bibinfo {volume}
  {11}},\ \bibinfo {pages} {983} (\bibinfo {year} {1978})}\BibitemShut
  {NoStop}%
\bibitem [{\citenamefont {Derrida}\ and\ \citenamefont
  {Spohn}(1988)}]{derrida1988polymers}%
  \BibitemOpen
  \bibfield  {author} {\bibinfo {author} {\bibfnamefont {B.}~\bibnamefont
  {Derrida}}\ and\ \bibinfo {author} {\bibfnamefont {H.}~\bibnamefont
  {Spohn}},\ }\href@noop {} {\bibfield  {journal} {\bibinfo  {journal} {Journal
  of Statistical Physics}\ }\textbf {\bibinfo {volume} {51}},\ \bibinfo {pages}
  {817} (\bibinfo {year} {1988})}\BibitemShut {NoStop}%
\bibitem [{\citenamefont {Ioffe}\ and\ \citenamefont
  {M{\'e}zard}(2010)}]{ioffe2010disorder}%
  \BibitemOpen
  \bibfield  {author} {\bibinfo {author} {\bibfnamefont {L.}~\bibnamefont
  {Ioffe}}\ and\ \bibinfo {author} {\bibfnamefont {M.}~\bibnamefont
  {M{\'e}zard}},\ }\href@noop {} {\bibfield  {journal} {\bibinfo  {journal}
  {Physical review letters}\ }\textbf {\bibinfo {volume} {105}},\ \bibinfo
  {pages} {037001} (\bibinfo {year} {2010})}\BibitemShut {NoStop}%
\bibitem [{\citenamefont {Pietracaprina}\ \emph {et~al.}(2016)\citenamefont
  {Pietracaprina}, \citenamefont {Ros},\ and\ \citenamefont
  {Scardicchio}}]{pietracaprina16forward}%
  \BibitemOpen
  \bibfield  {author} {\bibinfo {author} {\bibfnamefont {F.}~\bibnamefont
  {Pietracaprina}}, \bibinfo {author} {\bibfnamefont {V.}~\bibnamefont {Ros}},
  \ and\ \bibinfo {author} {\bibfnamefont {A.}~\bibnamefont {Scardicchio}},\
  }\href {\doibase 10.1103/PhysRevB.93.054201} {\bibfield  {journal} {\bibinfo
  {journal} {Phys. Rev. B}\ }\textbf {\bibinfo {volume} {93}},\ \bibinfo
  {pages} {054201} (\bibinfo {year} {2016})}\BibitemShut {NoStop}%
\bibitem [{\citenamefont {Barahona}(1982)}]{barahona}%
  \BibitemOpen
  \bibfield  {author} {\bibinfo {author} {\bibfnamefont {F.}~\bibnamefont
  {Barahona}},\ }\href {http://stacks.iop.org/0305-4470/15/i=10/a=028}
  {\bibfield  {journal} {\bibinfo  {journal} {Journal of Physics A:
  Mathematical and General}\ }\textbf {\bibinfo {volume} {15}},\ \bibinfo
  {pages} {3241} (\bibinfo {year} {1982})}\BibitemShut {NoStop}%
\bibitem [{\citenamefont {Pinsker}(1973)}]{Pinsker73}%
  \BibitemOpen
  \bibfield  {author} {\bibinfo {author} {\bibfnamefont {M.~S.}\ \bibnamefont
  {Pinsker}},\ }in\ \href@noop {} {\emph {\bibinfo {booktitle} {7th
  International Teletraffic Conference}}}\ (\bibinfo {year} {1973})\BibitemShut
  {NoStop}%
\bibitem [{\citenamefont {Zagoskin}\ \emph {et~al.}(2014)\citenamefont
  {Zagoskin}, \citenamefont {Il'ichev}, \citenamefont {Grajcar}, \citenamefont
  {Betouras},\ and\ \citenamefont {Nori}}]{zagoskin2014}%
  \BibitemOpen
  \bibfield  {author} {\bibinfo {author} {\bibfnamefont {A.~M.}\ \bibnamefont
  {Zagoskin}}, \bibinfo {author} {\bibfnamefont {E.}~\bibnamefont {Il'ichev}},
  \bibinfo {author} {\bibfnamefont {M.}~\bibnamefont {Grajcar}}, \bibinfo
  {author} {\bibfnamefont {J.~J.}\ \bibnamefont {Betouras}}, \ and\ \bibinfo
  {author} {\bibfnamefont {F.}~\bibnamefont {Nori}},\ }\href {\doibase
  10.3389/fphy.2014.00033} {\bibfield  {journal} {\bibinfo  {journal}
  {Frontiers in Physics}\ }\textbf {\bibinfo {volume} {2}} (\bibinfo {year}
  {2014}),\ 10.3389/fphy.2014.00033}\BibitemShut {NoStop}%
\bibitem [{\citenamefont {Bauer}\ \emph {et~al.}(2015)\citenamefont {Bauer},
  \citenamefont {Wang}, \citenamefont {Pižorn},\ and\ \citenamefont
  {Troyer}}]{bauer2015}%
  \BibitemOpen
  \bibfield  {author} {\bibinfo {author} {\bibfnamefont {B.}~\bibnamefont
  {Bauer}}, \bibinfo {author} {\bibfnamefont {L.}~\bibnamefont {Wang}},
  \bibinfo {author} {\bibfnamefont {I.}~\bibnamefont {Pižorn}}, \ and\
  \bibinfo {author} {\bibfnamefont {M.}~\bibnamefont {Troyer}},\ }\href@noop {}
  {\bibfield  {journal} {\bibinfo  {journal} {arXiv preprint
  arXiv:1501.06914v1}\ } (\bibinfo {year} {2015})}\BibitemShut {NoStop}%
\bibitem [{\citenamefont {Or\'us}\ and\ \citenamefont
  {Latorre}(2004)}]{Orus2004AdiabEntangl}%
  \BibitemOpen
  \bibfield  {author} {\bibinfo {author} {\bibfnamefont {R.}~\bibnamefont
  {Or\'us}}\ and\ \bibinfo {author} {\bibfnamefont {J.~I.}\ \bibnamefont
  {Latorre}},\ }\href {\doibase 10.1103/PhysRevA.69.052308} {\bibfield
  {journal} {\bibinfo  {journal} {Phys. Rev. A}\ }\textbf {\bibinfo {volume}
  {69}},\ \bibinfo {pages} {052308} (\bibinfo {year} {2004})}\BibitemShut
  {NoStop}%
\bibitem [{\citenamefont {Vidal}(2003)}]{Vidal2003EffClassSim}%
  \BibitemOpen
  \bibfield  {author} {\bibinfo {author} {\bibfnamefont {G.}~\bibnamefont
  {Vidal}},\ }\href {\doibase 10.1103/PhysRevLett.91.147902} {\bibfield
  {journal} {\bibinfo  {journal} {Phys. Rev. Lett.}\ }\textbf {\bibinfo
  {volume} {91}},\ \bibinfo {pages} {147902} (\bibinfo {year}
  {2003})}\BibitemShut {NoStop}%
\bibitem [{\citenamefont {Jozsa}\ and\ \citenamefont
  {Linden}(2003)}]{jozsa2011}%
  \BibitemOpen
  \bibfield  {author} {\bibinfo {author} {\bibfnamefont {R.}~\bibnamefont
  {Jozsa}}\ and\ \bibinfo {author} {\bibfnamefont {N.}~\bibnamefont {Linden}},\
  }\href {\doibase 10.1098/rspa.2002.1097} {\bibfield  {journal} {\bibinfo
  {journal} {Proceedings of the Royal Society of London A: Mathematical,
  Physical and Engineering Sciences}\ }\textbf {\bibinfo {volume} {459}},\
  \bibinfo {pages} {2011} (\bibinfo {year} {2003})}\BibitemShut {NoStop}%
\bibitem [{\citenamefont {Humeniuk}\ and\ \citenamefont
  {Roscilde}(2012)}]{roscilde}%
  \BibitemOpen
  \bibfield  {author} {\bibinfo {author} {\bibfnamefont {S.}~\bibnamefont
  {Humeniuk}}\ and\ \bibinfo {author} {\bibfnamefont {T.}~\bibnamefont
  {Roscilde}},\ }\href@noop {} {\bibfield  {journal} {\bibinfo  {journal}
  {Physical Review B}\ }\textbf {\bibinfo {volume} {86}},\ \bibinfo {pages}
  {235116} (\bibinfo {year} {2012})}\BibitemShut {NoStop}%
\bibitem [{\citenamefont {Steger}\ and\ \citenamefont
  {Wormald}(1999)}]{steger1999generate}%
  \BibitemOpen
  \bibfield  {author} {\bibinfo {author} {\bibfnamefont {A.}~\bibnamefont
  {Steger}}\ and\ \bibinfo {author} {\bibfnamefont {N.~C.}\ \bibnamefont
  {Wormald}},\ }\href {\doibase 10.1017/S0963548399003867} {\bibfield
  {journal} {\bibinfo  {journal} {Comb. Probab. Comput.}\ }\textbf {\bibinfo
  {volume} {8}},\ \bibinfo {pages} {377} (\bibinfo {year} {1999})}\BibitemShut
  {NoStop}%
\bibitem [{\citenamefont {Amico}\ \emph {et~al.}(2008)\citenamefont {Amico},
  \citenamefont {Fazio}, \citenamefont {Osterloh},\ and\ \citenamefont
  {Vedral}}]{Fazio2008review}%
  \BibitemOpen
  \bibfield  {author} {\bibinfo {author} {\bibfnamefont {L.}~\bibnamefont
  {Amico}}, \bibinfo {author} {\bibfnamefont {R.}~\bibnamefont {Fazio}},
  \bibinfo {author} {\bibfnamefont {A.}~\bibnamefont {Osterloh}}, \ and\
  \bibinfo {author} {\bibfnamefont {V.}~\bibnamefont {Vedral}},\ }\href
  {\doibase 10.1103/RevModPhys.80.517} {\bibfield  {journal} {\bibinfo
  {journal} {Rev. Mod. Phys.}\ }\textbf {\bibinfo {volume} {80}},\ \bibinfo
  {pages} {517} (\bibinfo {year} {2008})}\BibitemShut {NoStop}%
\bibitem [{\citenamefont {Hyllus}\ \emph {et~al.}(2012)\citenamefont {Hyllus},
  \citenamefont {Laskowski}, \citenamefont {Krischek}, \citenamefont
  {Schwemmer}, \citenamefont {Wieczorek}, \citenamefont {Weinfurter},
  \citenamefont {Pezz\'e},\ and\ \citenamefont {Smerzi}}]{hyllus2012}%
  \BibitemOpen
  \bibfield  {author} {\bibinfo {author} {\bibfnamefont {P.}~\bibnamefont
  {Hyllus}}, \bibinfo {author} {\bibfnamefont {W.}~\bibnamefont {Laskowski}},
  \bibinfo {author} {\bibfnamefont {R.}~\bibnamefont {Krischek}}, \bibinfo
  {author} {\bibfnamefont {C.}~\bibnamefont {Schwemmer}}, \bibinfo {author}
  {\bibfnamefont {W.}~\bibnamefont {Wieczorek}}, \bibinfo {author}
  {\bibfnamefont {H.}~\bibnamefont {Weinfurter}}, \bibinfo {author}
  {\bibfnamefont {L.}~\bibnamefont {Pezz\'e}}, \ and\ \bibinfo {author}
  {\bibfnamefont {A.}~\bibnamefont {Smerzi}},\ }\href {\doibase
  10.1103/PhysRevA.85.022321} {\bibfield  {journal} {\bibinfo  {journal} {Phys.
  Rev. A}\ }\textbf {\bibinfo {volume} {85}},\ \bibinfo {pages} {022321}
  (\bibinfo {year} {2012})}\BibitemShut {NoStop}%
\bibitem [{\citenamefont {Edwards}\ and\ \citenamefont
  {Anderson}(1975)}]{edwardsanderson1975}%
  \BibitemOpen
  \bibfield  {author} {\bibinfo {author} {\bibfnamefont {S.~F.}\ \bibnamefont
  {Edwards}}\ and\ \bibinfo {author} {\bibfnamefont {P.~W.}\ \bibnamefont
  {Anderson}},\ }\href {http://stacks.iop.org/0305-4608/5/i=5/a=017} {\bibfield
   {journal} {\bibinfo  {journal} {Journal of Physics F: Metal Physics}\
  }\textbf {\bibinfo {volume} {5}},\ \bibinfo {pages} {965} (\bibinfo {year}
  {1975})}\BibitemShut {NoStop}%
\bibitem [{\citenamefont {Metz}\ \emph {et~al.}(2014)\citenamefont {Metz},
  \citenamefont {Parisi},\ and\ \citenamefont {Leuzzi}}]{PhysRevE.90.052109}%
  \BibitemOpen
  \bibfield  {author} {\bibinfo {author} {\bibfnamefont {F.~L.}\ \bibnamefont
  {Metz}}, \bibinfo {author} {\bibfnamefont {G.}~\bibnamefont {Parisi}}, \ and\
  \bibinfo {author} {\bibfnamefont {L.}~\bibnamefont {Leuzzi}},\ }\href
  {\doibase 10.1103/PhysRevE.90.052109} {\bibfield  {journal} {\bibinfo
  {journal} {Phys. Rev. E}\ }\textbf {\bibinfo {volume} {90}},\ \bibinfo
  {pages} {052109} (\bibinfo {year} {2014})}\BibitemShut {NoStop}%
\bibitem [{\citenamefont {Ray}\ \emph {et~al.}(1989)\citenamefont {Ray},
  \citenamefont {Chakrabarti},\ and\ \citenamefont
  {Chakrabarti}}]{PhysRevB.39.11828}%
  \BibitemOpen
  \bibfield  {author} {\bibinfo {author} {\bibfnamefont {P.}~\bibnamefont
  {Ray}}, \bibinfo {author} {\bibfnamefont {B.~K.}\ \bibnamefont
  {Chakrabarti}}, \ and\ \bibinfo {author} {\bibfnamefont {A.}~\bibnamefont
  {Chakrabarti}},\ }\href {\doibase 10.1103/PhysRevB.39.11828} {\bibfield
  {journal} {\bibinfo  {journal} {Phys. Rev. B}\ }\textbf {\bibinfo {volume}
  {39}},\ \bibinfo {pages} {11828} (\bibinfo {year} {1989})}\BibitemShut
  {NoStop}%
\bibitem [{\citenamefont {Schiulaz}\ and\ \citenamefont
  {Müller}(2014)}]{schiulaz2013ideal}%
  \BibitemOpen
  \bibfield  {author} {\bibinfo {author} {\bibfnamefont {M.}~\bibnamefont
  {Schiulaz}}\ and\ \bibinfo {author} {\bibfnamefont {M.}~\bibnamefont
  {Müller}},\ }\href@noop {} {\bibfield  {journal} {\bibinfo  {journal} {AIP
  Conference Proceedings}\ }\textbf {\bibinfo {volume} {1610}} (\bibinfo {year}
  {2014})}\BibitemShut {NoStop}%
\bibitem [{\citenamefont {Pino}\ \emph {et~al.}(2016)\citenamefont {Pino},
  \citenamefont {Ioffe},\ and\ \citenamefont {Altshuler}}]{pino2015non}%
  \BibitemOpen
  \bibfield  {author} {\bibinfo {author} {\bibfnamefont {M.}~\bibnamefont
  {Pino}}, \bibinfo {author} {\bibfnamefont {L.~B.}\ \bibnamefont {Ioffe}}, \
  and\ \bibinfo {author} {\bibfnamefont {B.~L.}\ \bibnamefont {Altshuler}},\
  }\href {\doibase 10.1073/pnas.1520033113} {\bibfield  {journal} {\bibinfo
  {journal} {Proceedings of the National Academy of Sciences}\ }\textbf
  {\bibinfo {volume} {113}},\ \bibinfo {pages} {536} (\bibinfo {year}
  {2016})}\BibitemShut {NoStop}%
\bibitem [{\citenamefont {Yao}\ \emph {et~al.}(2014)\citenamefont {Yao},
  \citenamefont {Laumann}, \citenamefont {Cirac}, \citenamefont {Lukin},\ and\
  \citenamefont {Moore}}]{yao2014quasi}%
  \BibitemOpen
  \bibfield  {author} {\bibinfo {author} {\bibfnamefont {N.}~\bibnamefont
  {Yao}}, \bibinfo {author} {\bibfnamefont {C.}~\bibnamefont {Laumann}},
  \bibinfo {author} {\bibfnamefont {J.~I.}\ \bibnamefont {Cirac}}, \bibinfo
  {author} {\bibfnamefont {M.}~\bibnamefont {Lukin}}, \ and\ \bibinfo {author}
  {\bibfnamefont {J.}~\bibnamefont {Moore}},\ }\href@noop {} {\bibfield
  {journal} {\bibinfo  {journal} {arXiv preprint arXiv:1410.7407}\ } (\bibinfo
  {year} {2014})}\BibitemShut {NoStop}%
\bibitem [{\citenamefont {van Dam}\ \emph {et~al.}(2001)\citenamefont {van
  Dam}, \citenamefont {Mosca},\ and\ \citenamefont {Vazirani}}]{vanDam2001}%
  \BibitemOpen
  \bibfield  {author} {\bibinfo {author} {\bibfnamefont {W.}~\bibnamefont {van
  Dam}}, \bibinfo {author} {\bibfnamefont {M.}~\bibnamefont {Mosca}}, \ and\
  \bibinfo {author} {\bibfnamefont {U.}~\bibnamefont {Vazirani}},\ }in\ \href
  {\doibase 10.1109/SFCS.2001.959902} {\emph {\bibinfo {booktitle} {Foundations
  of Computer Science, 2001. Proceedings. 42nd IEEE Symposium on}}}\ (\bibinfo
  {year} {2001})\ pp.\ \bibinfo {pages} {279--287}\BibitemShut {NoStop}%
\bibitem [{\citenamefont {Suzuki}(1976)}]{suzuki1}%
  \BibitemOpen
  \bibfield  {author} {\bibinfo {author} {\bibfnamefont {M.}~\bibnamefont
  {Suzuki}},\ }\href@noop {} {\bibfield  {journal} {\bibinfo  {journal}
  {Progress of Theoretical Physics}\ }\textbf {\bibinfo {volume} {56}},\
  \bibinfo {pages} {1454} (\bibinfo {year} {1976})}\BibitemShut {NoStop}%
\bibitem [{\citenamefont {Suzuki}(2012)}]{suzuki2}%
  \BibitemOpen
  \bibfield  {author} {\bibinfo {author} {\bibfnamefont {M.}~\bibnamefont
  {Suzuki}},\ }\href@noop {} {\emph {\bibinfo {title} {Quantum Monte Carlo
  Methods in Equilibrium and Nonequilibrium Systems: Proceedings of the Ninth
  Taniguchi International Symposium, Susono, Japan, November 14--18, 1986}}},\
  Vol.~\bibinfo {volume} {74}\ (\bibinfo  {publisher} {Springer Science \&
  Business Media},\ \bibinfo {year} {2012})\BibitemShut {NoStop}%
\bibitem [{\citenamefont {Santoro}\ \emph {et~al.}(2002)\citenamefont
  {Santoro}, \citenamefont {Marto{\v{n}}{\'a}k}, \citenamefont {Tosatti},\ and\
  \citenamefont {Car}}]{car}%
  \BibitemOpen
  \bibfield  {author} {\bibinfo {author} {\bibfnamefont {G.~E.}\ \bibnamefont
  {Santoro}}, \bibinfo {author} {\bibfnamefont {R.}~\bibnamefont
  {Marto{\v{n}}{\'a}k}}, \bibinfo {author} {\bibfnamefont {E.}~\bibnamefont
  {Tosatti}}, \ and\ \bibinfo {author} {\bibfnamefont {R.}~\bibnamefont
  {Car}},\ }\href@noop {} {\bibfield  {journal} {\bibinfo  {journal} {Science}\
  }\textbf {\bibinfo {volume} {295}},\ \bibinfo {pages} {2427} (\bibinfo {year}
  {2002})}\BibitemShut {NoStop}%
\bibitem [{\citenamefont {Marto{\v{n}}{\'a}k}\ \emph
  {et~al.}(2002)\citenamefont {Marto{\v{n}}{\'a}k}, \citenamefont {Santoro},\
  and\ \citenamefont {Tosatti}}]{martovnak}%
  \BibitemOpen
  \bibfield  {author} {\bibinfo {author} {\bibfnamefont {R.}~\bibnamefont
  {Marto{\v{n}}{\'a}k}}, \bibinfo {author} {\bibfnamefont {G.~E.}\ \bibnamefont
  {Santoro}}, \ and\ \bibinfo {author} {\bibfnamefont {E.}~\bibnamefont
  {Tosatti}},\ }\href@noop {} {\bibfield  {journal} {\bibinfo  {journal}
  {Physical Review B}\ }\textbf {\bibinfo {volume} {66}},\ \bibinfo {pages}
  {094203} (\bibinfo {year} {2002})}\BibitemShut {NoStop}%
\bibitem [{\citenamefont {Young}\ \emph {et~al.}(2008)\citenamefont {Young},
  \citenamefont {Knysh},\ and\ \citenamefont {Smelyanskiy}}]{young}%
  \BibitemOpen
  \bibfield  {author} {\bibinfo {author} {\bibfnamefont {A.~P.}\ \bibnamefont
  {Young}}, \bibinfo {author} {\bibfnamefont {S.}~\bibnamefont {Knysh}}, \ and\
  \bibinfo {author} {\bibfnamefont {V.~N.}\ \bibnamefont {Smelyanskiy}},\
  }\href@noop {} {\bibfield  {journal} {\bibinfo  {journal} {Physical review
  letters}\ }\textbf {\bibinfo {volume} {101}},\ \bibinfo {pages} {170503}
  (\bibinfo {year} {2008})}\BibitemShut {NoStop}%
\bibitem [{\citenamefont {Melko}\ \emph {et~al.}(2010)\citenamefont {Melko},
  \citenamefont {Kallin},\ and\ \citenamefont {Hastings}}]{melko}%
  \BibitemOpen
  \bibfield  {author} {\bibinfo {author} {\bibfnamefont {R.~G.}\ \bibnamefont
  {Melko}}, \bibinfo {author} {\bibfnamefont {A.~B.}\ \bibnamefont {Kallin}}, \
  and\ \bibinfo {author} {\bibfnamefont {M.~B.}\ \bibnamefont {Hastings}},\
  }\href {\doibase 10.1103/PhysRevB.82.100409} {\bibfield  {journal} {\bibinfo
  {journal} {Phys. Rev. B}\ }\textbf {\bibinfo {volume} {82}},\ \bibinfo
  {pages} {100409} (\bibinfo {year} {2010})}\BibitemShut {NoStop}%
\bibitem [{\citenamefont {Hastings}\ \emph {et~al.}(2010)\citenamefont
  {Hastings}, \citenamefont {Gonz{\'a}lez}, \citenamefont {Kallin},\ and\
  \citenamefont {Melko}}]{hastings}%
  \BibitemOpen
  \bibfield  {author} {\bibinfo {author} {\bibfnamefont {M.~B.}\ \bibnamefont
  {Hastings}}, \bibinfo {author} {\bibfnamefont {I.}~\bibnamefont
  {Gonz{\'a}lez}}, \bibinfo {author} {\bibfnamefont {A.~B.}\ \bibnamefont
  {Kallin}}, \ and\ \bibinfo {author} {\bibfnamefont {R.~G.}\ \bibnamefont
  {Melko}},\ }\href@noop {} {\bibfield  {journal} {\bibinfo  {journal}
  {Physical review letters}\ }\textbf {\bibinfo {volume} {104}},\ \bibinfo
  {pages} {157201} (\bibinfo {year} {2010})}\BibitemShut {NoStop}%
\bibitem [{\citenamefont {Caraglio}\ and\ \citenamefont
  {Gliozzi}(2008)}]{caraglio}%
  \BibitemOpen
  \bibfield  {author} {\bibinfo {author} {\bibfnamefont {M.}~\bibnamefont
  {Caraglio}}\ and\ \bibinfo {author} {\bibfnamefont {F.}~\bibnamefont
  {Gliozzi}},\ }\href@noop {} {\bibfield  {journal} {\bibinfo  {journal}
  {Journal of High Energy Physics}\ }\textbf {\bibinfo {volume} {2008}},\
  \bibinfo {pages} {076} (\bibinfo {year} {2008})}\BibitemShut {NoStop}%
\bibitem [{\citenamefont {Gliozzi}\ and\ \citenamefont
  {Tagliacozzo}(2010)}]{gliozzi}%
  \BibitemOpen
  \bibfield  {author} {\bibinfo {author} {\bibfnamefont {F.}~\bibnamefont
  {Gliozzi}}\ and\ \bibinfo {author} {\bibfnamefont {L.}~\bibnamefont
  {Tagliacozzo}},\ }\href@noop {} {\bibfield  {journal} {\bibinfo  {journal}
  {Journal of Statistical Mechanics: Theory and Experiment}\ }\textbf {\bibinfo
  {volume} {2010}},\ \bibinfo {pages} {P01002} (\bibinfo {year}
  {2010})}\BibitemShut {NoStop}%
\bibitem [{\citenamefont {Alba}\ \emph {et~al.}(2011)\citenamefont {Alba},
  \citenamefont {Tagliacozzo},\ and\ \citenamefont {Calabrese}}]{alba}%
  \BibitemOpen
  \bibfield  {author} {\bibinfo {author} {\bibfnamefont {V.}~\bibnamefont
  {Alba}}, \bibinfo {author} {\bibfnamefont {L.}~\bibnamefont {Tagliacozzo}}, \
  and\ \bibinfo {author} {\bibfnamefont {P.}~\bibnamefont {Calabrese}},\
  }\href@noop {} {\bibfield  {journal} {\bibinfo  {journal} {Journal of
  Statistical Mechanics: Theory and Experiment}\ }\textbf {\bibinfo {volume}
  {2011}},\ \bibinfo {pages} {P06012} (\bibinfo {year} {2011})}\BibitemShut
  {NoStop}%
\bibitem [{\citenamefont {Buividovich}\ and\ \citenamefont
  {Polikarpov}(2008)}]{buividovich}%
  \BibitemOpen
  \bibfield  {author} {\bibinfo {author} {\bibfnamefont {P.}~\bibnamefont
  {Buividovich}}\ and\ \bibinfo {author} {\bibfnamefont {M.}~\bibnamefont
  {Polikarpov}},\ }\href@noop {} {\bibfield  {journal} {\bibinfo  {journal}
  {Nuclear Physics B}\ }\textbf {\bibinfo {volume} {802}},\ \bibinfo {pages}
  {458 } (\bibinfo {year} {2008})}\BibitemShut {NoStop}%
\bibitem [{\citenamefont {Calabrese}\ and\ \citenamefont
  {Cardy}(2004)}]{calabrese}%
  \BibitemOpen
  \bibfield  {author} {\bibinfo {author} {\bibfnamefont {P.}~\bibnamefont
  {Calabrese}}\ and\ \bibinfo {author} {\bibfnamefont {J.}~\bibnamefont
  {Cardy}},\ }\href@noop {} {\bibfield  {journal} {\bibinfo  {journal} {Journal
  of Statistical Mechanics: Theory and Experiment}\ }\textbf {\bibinfo {volume}
  {2004}},\ \bibinfo {pages} {P06002} (\bibinfo {year} {2004})}\BibitemShut
  {NoStop}%
\bibitem [{\citenamefont {Prokof'ev}\ \emph {et~al.}(1998)\citenamefont
  {Prokof'ev}, \citenamefont {Svistunov},\ and\ \citenamefont
  {Tupitsyn}}]{tupitsyn}%
  \BibitemOpen
  \bibfield  {author} {\bibinfo {author} {\bibfnamefont {N.}~\bibnamefont
  {Prokof'ev}}, \bibinfo {author} {\bibfnamefont {B.}~\bibnamefont
  {Svistunov}}, \ and\ \bibinfo {author} {\bibfnamefont {I.}~\bibnamefont
  {Tupitsyn}},\ }\href@noop {} {\bibfield  {journal} {\bibinfo  {journal}
  {Physics Letters A}\ }\textbf {\bibinfo {volume} {238}},\ \bibinfo {pages}
  {253} (\bibinfo {year} {1998})}\BibitemShut {NoStop}%
\bibitem [{\citenamefont {Fr\"owis}\ and\ \citenamefont
  {D\"ur}(2012)}]{macroscopicity}%
  \BibitemOpen
  \bibfield  {author} {\bibinfo {author} {\bibfnamefont {F.}~\bibnamefont
  {Fr\"owis}}\ and\ \bibinfo {author} {\bibfnamefont {W.}~\bibnamefont
  {D\"ur}},\ }\href {http://stacks.iop.org/1367-2630/14/i=9/a=093039}
  {\bibfield  {journal} {\bibinfo  {journal} {New Journal of Physics}\ }\textbf
  {\bibinfo {volume} {14}},\ \bibinfo {pages} {093039} (\bibinfo {year}
  {2012})}\BibitemShut {NoStop}%
\end{thebibliography}
\end{document}